\documentstyle[12pt,epsfig,rotating] {article}
\begin{document}
\newcommand{\sqrts}{\mbox{$\protect \sqrt{s}$}}
\newcommand{\sqrtsp}{\mbox{$\sqrt{s'}$}}
\newcommand{\epm} {\mbox{$\mathrm{e}^+ \mathrm{e}^-$}}
\newcommand{\mpm} {\mbox{$\mu^+ \mu^-$}}
\newcommand{\nprode}{\mbox{$N^{\mathrm{e}^+ \mathrm{e}^-}_{prod}$}}
\newcommand{\nprodm}{\mbox{$N^{\mu^+ \mu^-}_{prod}$}}
\newcommand{\nexpe}{\mbox{$N_{exp}^{\mathrm{e}^+ \mathrm{e}^-}$}}
\newcommand{\nexpm}{\mbox{$N_{exp}^{\mu^+ \mu^-}$}}
\newcommand{\nexpt}{\mbox{$N_{exp}^{total}$}}
\newcommand{\Zo}{\mbox{$\protect {\rm Z}^0$}}
\newcommand{\bZo}{{\bf \mbox{$\rm Z}^0$}}
\newcommand{\Zs}{\mbox{$\mathrm{Z}^{*}$}}
\newcommand{\ho}{\mbox{$\mathrm{h}^{0}$}}
\newcommand{\Ho}{\mbox{$\mathrm{H}^{0}$}}
\newcommand{\Ao}{\mbox{$\mathrm{A}^{0}$}}
\newcommand{\Wpm}{\mbox{$\mathrm{W}^{\pm}$}}
\newcommand{\Hpm}{\mbox{$\mathrm{H}^{\pm}$}}
\newcommand{\WW}{\mbox{$\mathrm{W}^{+}\mathrm{W}^{-}$}}
\newcommand{\ZZ}{\mbox{$\mathrm{Z}^{0}\mathrm{Z}^{0}$}}
\newcommand{\ko}{\mbox{${\tilde{\chi_1}^0}$}}
\newcommand{\koko}{\mbox{${\tilde{\chi_1}^0}{{\tilde{\chi_1}^0}}$}}
\newcommand{\bho}{\mbox{$\boldmath{\mathrm{H}^{0}}$}}
\newcommand{\ee}{\mbox{$\mathrm{e}^{+}\mathrm{e}^{-}$}}
\newcommand{\bee}{\mbox{$\boldmath {\mathrm{e}^{+}\mathrm{e}^{-}} $}}
\newcommand{\mm}{\mbox{$\mu^{+}\mu^{-}$}}
\newcommand{\bmm}{\mbox{$\boldmath {\mu^{+}\mu^{-}} $}}
\newcommand{\nn}{\mbox{$\nu \bar{\nu}$}}
\newcommand{\bnn}{\mbox{$\boldmath {\nu \bar{\nu}} $}}
\newcommand{\qq}{\mbox{$\protect {\rm q} \protect \bar{{\rm q}}$}}
\newcommand{\ff}{\mbox{$\mathrm{f} \bar{\mathrm{f}}$}}
\newcommand{\bqq}{\mbox{$\boldmath {\mathrm{q} \bar{\mathrm{q}}} $}}
\newcommand{\pb}{\mbox{$\protect {\rm pb}^{-1}$}}
\newcommand{\ra}{\mbox{$\rightarrow$}}
\newcommand{\br}{\mbox{$\boldmath {\rightarrow}$}}
\newcommand{\erh}{\mbox{$\mathrm{e}^+\mathrm{e}^-\rightarrow\mathrm{hadrons}$}}
\newcommand{\tptm}{\mbox{$\tau^{+}\tau^{-}$}}
\newcommand{\tpm}{\mbox{$\tau^{\pm}$}}
\newcommand{\pzvis}{\mbox{$\protect P^z_{\rm vis}$}}
\newcommand{\evisn}{\mbox{$\protect E_{\rm vis}$/$\protect \sqrt{s}$}}

\newcommand{\gamgam}{\mbox{$\gamma\gamma$}}
\newcommand{\uu}{\mbox{$\mathrm{u} \bar{\mathrm{u}}$}}
\newcommand{\dd}{\mbox{$\mathrm{d} \bar{\mathrm{d}}$}}
\newcommand{\bb}{\mbox{$\mathrm{b} \bar{\mathrm{b}}$}}
\newcommand{\cc}{\mbox{$\mathrm{c} \bar{\mathrm{c}}$}}
\newcommand{\nunu}{\mbox{$\nu \bar{\nu}$}}
\newcommand{\mZ}{\mbox{$m_{\mathrm{Z}^{0}}$}}
\newcommand{\mH}{\mbox{$m_{\mathrm{H}^{0}}$}}
\newcommand{\mh}{\mbox{$m_{\mathrm{h}^{0}}$}}
\newcommand{\mA}{\mbox{$m_{\mathrm{A}^{0}}$}}
\newcommand{\mHpm}{\mbox{$m_{\mathrm{H}^{\pm}}$}}
\newcommand{\mW}{\mbox{$m_{\mathrm{W}^{\pm}}$}}
\newcommand{\mtop}{\mbox{$m_{\mathrm{t}}$}}
\newcommand{\mb}{\mbox{$m_{\mathrm{b}}$}}
\newcommand{\lpm}{\mbox{$\ell ^+ \ell^-$}}
\newcommand{\G}{\mbox{$\mathrm{GeV}$}}
\newcommand{\Gc}{\mbox{$\mathrm{GeV}$}}
\newcommand{\Gcs}{\mbox{$\mathrm{GeV}$}}
\newcommand{\epsnn}{\mbox{$\epsilon^{\nu\bar{\nu}}$(\%)}}
\newcommand{\Nnn}{\mbox{$N^{\nu \bar{\nu}}_{exp}$}}
\newcommand{\epsll}{\mbox{$\epsilon^{\ell^{+}\ell^{-}}$(\%)}}
\newcommand{\Nll}{\mbox{$N^{\ell^+\ell^-}_{exp}$}}
\newcommand{\Nexp}{\mbox{$N^{total}_{exp}$}}
\newcommand{\kl}{\mbox{$\mathrm{K_{L}}$}}
\newcommand{\dedx}{\mbox{d$E$/d$x$}}
\newcommand{\etal}{\mbox{$et$ $al.$}}
\newcommand{\ie}{\mbox{$i.e.$}}
\newcommand{\sba}{\mbox{$\sin ^2 (\beta -\alpha)$}}
\newcommand{\cba}{\mbox{$\cos ^2 (\beta -\alpha)$}}
\newcommand{\tanb}{\mbox{$\tan \beta$}}
\newcommand{\PhysLett}  {Phys.~Lett.}
\newcommand{\PRL} {Phys.~Rev.\ Lett.}
\newcommand{\PhysRep}   {Phys.~Rep.}
\newcommand{\PhysRev}   {Phys.~Rev.}
\newcommand{\NPhys}  {Nucl.~Phys.}
\newcommand{\NIM} {Nucl.~Instr.\ Meth.}
\newcommand{\CPC} {Comp.~Phys.\ Comm.}
\newcommand{\ZPhys}  {Z.~Phys.}
\newcommand{\IEEENS} {IEEE Trans.\ Nucl.~Sci.}

\begin{titlepage}
\begin{center}{\large   EUROPEAN LABORATORY FOR PARTICLE PHYSICS}
\end{center}
\begin{flushright}
      CERN-PPE/97-115  \\ 19 August 1997
\end{flushright}
\bigskip\bigskip

\begin{center}{\LARGE\bf   Search for the Standard Model Higgs Boson \\
in {\boldmath \ee\ \unboldmath} Collisions at 
$\sqrt s =$~161\ -\ 172~GeV }.
\end{center}
 \bigskip
\begin{center}{\LARGE The OPAL Collaboration}
\end{center}
%
\begin{center}{\large  Abstract}\end{center}
This paper describes a search for the Standard Model Higgs boson
using data from \ee\ collisions
collected at center-of-mass
energies of 161, 170 and  172~\Gc\  by the OPAL detector at LEP.
The data collected at these energies correspond to integrated
luminosities of 10.0, 1.0 and 9.4~\pb, respectively. The search is sensitive to 
the main final states from the
process in which the Higgs boson is produced in association with a 
fermion anti-fermion pair, namely  
four jets, two jets with missing energy, and 
two jets produced together with a pair of electron,
muon or tau leptons.
One candidate event is observed,
in agreement with the Standard Model background expectation.
In combination with previous OPAL
searches at center-of-mass energies close to the \Zo\ resonance
and the revised previous OPAL searches at
 161~\Gc\ , we derive a lower limit of 69.4~\Gc\ 
for the mass of the Standard Model Higgs boson
at the 95\% confidence level.
\begin{center}
{\large  (Submitted to Zeit. Phys. C)} 
\end{center}


\end{titlepage}

\begin{center}{\Large        The OPAL Collaboration
}\end{center}\bigskip
\begin{center}{
K.\thinspace Ackerstaff$^{  8}$,
G.\thinspace Alexander$^{ 23}$,
J.\thinspace Allison$^{ 16}$,
N.\thinspace Altekamp$^{  5}$,
K.J.\thinspace Anderson$^{  9}$,
S.\thinspace Anderson$^{ 12}$,
S.\thinspace Arcelli$^{  2}$,
S.\thinspace Asai$^{ 24}$,
D.\thinspace Axen$^{ 29}$,
G.\thinspace Azuelos$^{ 18,  a}$
% ...
}\end{center}\bigskip
\bigskip
$^{  1}$School of Physics and Space Research, University of Birmingham,
Birmingham B15 2TT, UK
\newline
$^{  2}$Dipartimento di Fisica dell' Universit\`a di Bologna and INFN,
I-40126 Bologna, Italy
\newline
% ...
$^{  5}$Cavendish Laboratory, Cambridge CB3 0HE, UK
\newline
% ...
$^{  8}$CERN, European Organisation for Particle Physics,
CH-1211 Geneva 23, Switzerland
\newline
$^{  9}$Enrico Fermi Institute and Department of Physics,
University of Chicago, Chicago IL 60637, USA
\newline
% ...
$^{ 12}$Indiana University, Department of Physics,
Swain Hall West 117, Bloomington IN 47405, USA
\newline
% ...
$^{ 16}$Department of Physics, Schuster Laboratory, The University,
Manchester M13 9PL, UK
\newline
% ...
$^{ 18}$Laboratoire de Physique Nucl\'eaire, Universit\'e de Montr\'eal,
Montr\'eal, Quebec H3C 3J7, Canada
\newline
% ...
$^{ 23}$Department of Physics and Astronomy, Tel Aviv University,
Tel Aviv 69978, Israel
\newline
$^{ 24}$International Centre for Elementary Particle Physics and
Department of Physics, University of Tokyo, Tokyo 113, and
Kobe University, Kobe 657, Japan
\newline
% ...
$^{ 29}$University of British Columbia, Department of Physics,
Vancouver BC V6T 1Z1, Canada
\newline
% ...
$^{  a}$ and at TRIUMF, Vancouver, Canada V6T 2A3
\newline
$^{  b}$ and Royal Society University Research Fellow
\newline
$^{  c}$ and Institute of Nuclear Research, Debrecen, Hungary
\newline
$^{  d}$ and Department of Experimental Physics, Lajos Kossuth
University, Debrecen, Hungary
\newline
$^{  e}$ and Department of Physics, New York University, NY 1003, USA
\newline
\bigskip
\newpage

\section{Introduction}
Locally gauge-invariant theories of the electroweak interaction introduce
spontaneous symmetry breaking to allow some of the gauge
bosons to acquire mass while keeping the theory renormalizable.
The Standard Model (SM) \cite{sm} is the simplest such theory and uses the
self-interaction of a single doublet of complex scalar fields 
\cite{higgs} to produce spontaneous symmetry breaking. This model
predicts the existence of one physical scalar particle, the Higgs
boson, \Ho, whose couplings are fixed but whose mass is  
not predicted.

 Despite a wide experimental effort,
the Higgs boson has not yet been discovered. The  
experimental lower limits
for its mass, \mH, obtained from large samples of
\Zo\ boson decays, are published in 
\cite{sm-all}. Recently the OPAL Collaboration has
updated its result by including the data collected during summer 1996 at a 
center-of-mass energy of \sqrts=161~\Gc\,   yielding a lower
limit of 68.5~\Gc\  on the Higgs boson mass \cite{opal-161}.
During the autumn of 1996 the center-of-mass energy of the LEP \ee\ collider 
was upgraded to 172~\Gc\@. 

At these  center-of-mass energies, the main production
process
for the SM Higgs boson is \ee\ra\Zo\Ho. 
The dominant decay is
\Ho\ra\bb, with a branching ratio of approximately 86\%.
Other relevant decay modes are:
\Ho\ra\tptm\ (8\%), \Ho\ra\cc\ (4\%), and
\Ho\ra gluons (2\%) \cite{spira}. For Higgs boson masses of current
interest, these branching ratios exhibit only a mild dependence on the 
Higgs boson mass. 

The searches described address the principal final state
topologies, which account for about 95\% of all Higgs boson final states,
namely: (i) the four jets channel, \ee\ra\Zo\Ho\ra\qq\bb;
(ii) the missing energy channel, mainly from  
\ee\ra\Zo\Ho\ra\nn\bb, with a small
contribution from the \WW\ fusion process \ee\ra\nn\Ho;
(iii) the tau channels, \ee\ra\Zo\Ho\ra\tptm\qq\ and \qq\tptm;
(iv) the muon and electron channels, 
predominantly from  \ee\ra\Zo\Ho\ra\mm\qq\ and \ee\qq,
with the latter including a small
contribution from the \ZZ\ fusion process \ee\ra\ee\Ho.

The present paper describes
the analysis of the data collected at 
170 and 172~\Gc\  energies,
an improved analysis of the four jets channel 
for the 161~\Gc\  data
based on likelihood rather than cut-based methods,
and the derivation of a new mass limit.

\section{Detector, Data, and Simulations}
This analysis uses 1.0~\pb\ of data recorded with
the OPAL detector \cite{detector}  at \sqrts=170~\Gc\@,  9.4~\pb\ at
\sqrts=172~\Gc\ 
and 
 10.0 ~\pb\  at
\sqrts=161~\Gc\    for the four jets channel.

OPAL is a multipurpose apparatus
with nearly complete solid angle coverage and excellent hermeticity.
The central tracking detector consists of two layers of
silicon microstrip detectors \cite{simvtx} with polar angle\footnote{OPAL uses a right-handed
coordinate system where the $+z$ direction is along the electron beam and
where $+x$ points to the center of the LEP ring.  
The polar angle, $\theta$, is
defined with respect to the $+z$ direction and the azimuthal angle, $\phi$,
with respect to the horizontal, $+x$ direction.} coverage
$|\cos\theta|<0.9$, immediately
outside the beam-pipe, followed by a high-precision vertex drift chamber,
a large-volume jet chamber, and $z$-chambers,
all in a uniform  
0.435~T axial magnetic field. A lead-glass electromagnetic calorimeter
is located outside the magnet coil, which, in combination with
the forward calorimeter, gamma catcher, and silicon-tungsten
luminometer \cite{sw}, provide acceptance 
down to 24 mrad from the beam direction.  The silicon-tungsten luminometer
serves to measure the integrated luminosity using small-angle Bhabha
scattering events \cite{lumino}.
The magnet return yoke is instrumented with streamer tubes
for hadron calorimetry and
is surrounded by several layers of muon chambers.
Events are reconstructed from
charged-particle tracks and
energy deposits (``clusters") in the electromagnetic and hadronic calorimeters.
The tracks and clusters are required to pass a set of quality requirements
similar to those used in
previous Higgs boson searches \cite{higgsold}.
In calculating the total visible energy and momentum, $E_{\rm vis}$
and $\vec{P}_{\rm vis}$, of events and of
individual jets, corrections are applied that reduce the effect of 
double-counting of energy in the case of tracks and clusters associated
to them.

The signal detection efficiencies and accepted background cross sections
are estimated using a variety of Monte Carlo samples
all processed through a full simulation \cite{gopal} of the OPAL detector. 
The HZHA generator \cite{hzha}, including initial-state radiation effects, 
is used to simulate Higgs boson
production processes. 
The generated partons are hadronized using JETSET
\cite{pythia}. Signal samples
are produced for fixed values of \mH\ between 40 \Gc\  and 80 \Gc\@.
The estimates of the different background processes are based primarily
on the following event generators:
PYTHIA~\cite{pythia} (\Zo$/\gamma^*$\ra\qq($\gamma$)), 
grc4f~\cite{grc4f} (four-fermion processes including \WW\ and 
\Zo${\rm Z}^*$),  
BHWIDE~\cite{bhwide} (\ee$(\gamma)$),
KORALZ~\cite{koralz} (\mm$(\gamma)$ and \tptm$(\gamma)$), PYTHIA,
PHOJET~\cite{phojet} and
Vermaseren~\cite{vermaseren} (hadronic and leptonic two-photon processes).

\section{The Four Jets Channel}
%
The process \ee\ra\Zo\Ho\ra\qq\bb\ amounts to approximately 60\% of the
Higgs boson signal topologies. It is characterized by four energetic hadronic
jets, large visible energy and signals from b-hadron decays. 
The backgrounds are \Zo$/\gamma^*$\ra\qq\ with and without initial state 
radiation  accompanied by hard gluon emission as well as
four-fermion processes,
in particular \ee\ra\WW.
The suppression of these backgrounds relies on the
kinematic reconstruction of the \Zo\ boson and on
the identification of b-quarks from the Higgs boson decay.
The tagging of particles containing b quarks proceeds by detecting 
displaced secondary vertices in three dimensions exploiting the 
high-resolution obtained from OPAL's silicon microvertex detector and using 
leptons with high transverse momenta with respect to the
jets to which they are assigned.

The selection of candidate events and the suppression of background is
done in two steps. A preselection using cuts
is first performed in order to retain  
only events which have some similarity to the
signal. The remaining events are then
analyzed using a  likelihood technique.

This channel was already analyzed and published
using cut-based methods for the 161~\Gc\@  data
\cite{opal-161}, however, using likelihood methods and a better
b-tag improves the signal efficiency significantly with the same
expected background.
The analysis of this channel was therefore repeated for the
161~\Gc\@  data with the new likelihood method.  

The analysis proceeds in the following way:

First, large parts of the \Zo$/\gamma^*$\ra\qq\ background are eliminated by selecting
well defined four jet topologies using the cuts listed below: 

\begin{itemize}
\item[(1)]
      The events must qualify as a hadronic final state
      according to ref.~\cite{tkmh}.
\item[(2)]
      The radiative process \ee\ra\Zo$\gamma$\ra\qq$\gamma$ is largely
      eliminated by requiring that the effective center-of-mass energy,
      \sqrtsp, obtained by discarding the radiative photon
      from the event \cite{sprim}, is at least 140~\Gc\@  (150~\Gc\@) at
      $ \sqrt s = 161$~\Gc\@ ($\sqrt s = 170-172$~\Gc\@).
\item[(3)]
      The final state particles of an event are grouped into four jets using
      the Durham algorithm~\cite{durham}. The jet resolution parameter, 
      $y_{34}$, at which the number of jets changes from 3 to 4, is required to
      be larger than 0.005. 
\item[(4)] 
      The \Zo$/\gamma^*$\ra\qq\ background is further suppressed by
      requiring that the event shape parameter $C$ \cite{cpar}, which
      is large for spherical events, is larger than 0.45.
\item[(5)] All four jets are required to contain at least two tracks
      and two electromagnetic calorimeter clusters.
\item[(6)] To discriminate against poorly reconstructed events, a kinematic
      fit, using energy and momentum conservation constraints (4C-fit), 
      is required to converge with a probability larger than 0.01.

      The \ee\ra\Ho\Zo\  hypothesis is tested by a kinematic fit which,
      in addition to the energy and momentum conservation constraints, also
      forces two of the four jets 
       to 
      have an invariant mass equal to 
      the \Zo~boson mass (5C-fit). 
      This fit is applied in turn to all six possible
      associations of the four jets to the \Zo~and \Ho~bosons. The fit
      is required to converge for at least one combination with a probability
      of at least 0.01. The combination 
      yielding the highest $\chi^2$-probability is selected.      
\end{itemize}

Table \ref{fourj_t1} shows the number of events selected in data and Monte 
Carlo at each stage of the cuts for both center-of-mass energies.
 After removing most of the two-photon background (cut(1)),
a good agreement between the observed data and the expected background
can be seen.
\begin{table}[htbp]
\begin{center}
\begin{tabular}{|l||r||r||r|r||c|} \hline 
 Cut& Data~~~~~~    & Total bkg.& \qq($\gamma$) &4-ferm. &Efficiency (\%) \\
    & 161 \Gc\  &           &               &        & \mH~= 65 \Gc\     \\
\hline \hline
(1) & 1572 & 1389.4 & 1345.9 &   52 & 99.9  \\
(2) &  395 &  377.5 &  351.1 &   26.4 & 89.5  \\
(3) &   65 &   54.1 &   38.0 &   16.1 & 81.4  \\
(4) &   51 &   40.6 &   26.2 &   14.4 & 75.6  \\
(5) &   49 &   33.2 &   21.5 &   11.7 & 70.5  \\
(6) &   26 &   24.2 &   14.0 &   10.2 & 62.3  \\ \hline \hline
${\cal L}^{HZ} > \mbox{0.9} $ 
    &    0 & 0.75 $ \pm $ 0.08  & 0.49 & 0.26  & 32.1 \\
\hline 
\multicolumn{6}{c}{ } \\ \hline 
 Cut& Data~~~~~~ & Total bkg.& \qq($\gamma$) &4-ferm. &Efficiency (\%) \\
    & 170-172 \Gc\  &           &               &        & \mH~= 68 \Gc\     \\
\hline 
(1) & 1409 & 1306.7 & 1189.7 &  117.0 & 99.7  \\
(2) &  367 &  381.2 &  312.8 &   68.4 & 87.8  \\
(3) &   93 &   84.7 &   33.5 &   51.2 & 79.3  \\
(4) &   77 &   70.9 &   22.9 &   48.0 & 75.8  \\
(5) &   68 &   60.3 &   18.1 &   42.2 & 70.2  \\
(6) &   56 &   50.3 &   12.6 &   37.7 & 61.4  \\ \hline \hline
${\cal L}^{HZ} > \mbox{0.925} $ 
    &    1 & 0.88 $ \pm $ 0.07 & 0.34 & 0.55 & 28.4 \\
\hline 
\end{tabular}
\end{center}
\caption[sig]{\sl The number of events after each cut of the selection
        for the data at $\sqrt{s}$ = 161  (170-172) \Gc\, 
        and the expected background in the four jets channel.
        The background estimates are normalized to 10.0  (10.4 \pb). 
        The last column shows the selection efficiencies for the
        \Zo\Ho\ra\qq\bb\ final state.}
\label{fourj_t1}
\label{fourj_t2}
\end{table}

Next, a likelihood technique is employed
in order  to classify the remaining events 
as either \Zo$/\gamma^*$\ra\qq\ (1), a four-fermion process (2) 
or \Zo\Ho\ra\qq\bb\ (3).  
To select signal events with low background, 
the information from 14 quantities (described below)
which provide a good separation 
between the three different event classes is combined. 
Half of the variables  are related to the kinematics of the events, and the
other half are related to b-tagging.\
For each event the measured values {$x_i$} (i=1...14)
of these variables are compared with probability density functions normalized
to unity obtained from Monte Carlo events processed through the full detector 
simulation. The likelihood for each event to belong to any of the three 
event classes is calculated as follows.
For a single variable, the probability for an event to belong to class $j$ is 
given by 
\begin{eqnarray*}
    p_i^j(x_i) = \frac{f_i^j(x_i)}{\sum_{k=1}^{3} f_i^k(x_i)},
\end{eqnarray*}
where $f_i^j(x_i)$ denotes the probability density for event class $j$ 
and variable $i$.
The likelihood function for class $j$ is defined as the normalized product of
the $p_i^j(x_i)$
\begin{eqnarray*}
{\cal L}^j(\vec{x}) = \frac{\prod_{i=1}^{n} p_i^j(x_i)}
              {\sum_{k=1}^{3} \prod_{i=1}^{n} p_i^k(x_i) },
\end{eqnarray*}
where $n=14$ is the number of variables.
In order to select an event as a candidate event
 its likelihood for being a signal event is required to be larger than
a certain value depending on the preferred signal/background 
ratio.

The first set of variables entering the likelihood function
exploit the different kinematics
of the background and signal events:
the smallest angle between any pair of jets (1), the smallest dijet mass (2), the highest
and lowest jet energy (3-4) and the dijet mass closest to \mZ\ (5). All these
quantities are calculated after the 4C-fit.  The probability
of the best kinematic fit requiring energy and momentum conservation and 
equal dijet masses (6) and the larger $\beta = p/E$ factor of the two dijet 
momenta calculated from the jet pair combination closest to the WW hypothesis (7)
are also used. The best jet pair combination under 
the WW hypothesis is determined by
minimizing $ (p_{\rm dijet1}-x)^2 + (p_{\rm dijet2}-x)^2$, where
$ x$ is 30 GeV at 170-172 GeV and 6 GeV at 161 GeV center
of mass energy.
This quantity vanishes for perfectly reconstructed on-shell WW events.

The second set of variables is used to tag  b-flavored  hadrons.
Secondary vertices are identified in each jet separately using the three
dimensional extension of the  method described in \cite{btag2}.
To improve the quality of the vertices,
each accepted vertex is required to have at least two tracks containing 2 silicon microvertex hits in both $r\phi$ and $z$.
The variables to discriminate between b flavor and lighter 
flavors are: sum of the decay length significances 
 of the vertices found in 
the candidate Higgs jets (8)\footnote{The decay length significance
 is defined as
$L/\sigma_L$, where $L$ is the three dimensional distance between
the primary and the secondary vertex
and $\sigma_L$ is the corresponding error.},
 the maximum number of tracks with significantly large
impact parameter (9)\footnote{Significant tracks are defined by 
$ d/\sigma_d > 2.5$
where $d$ denotes the two dimensional impact parameter of the track 
and $ \sigma_d$ the error on the impact parameter.}, 
 the invariant
mass of the tracks in the
vertex with the largest decay length
significance (10),  
the smaller of the two Higgs jet masses (11), the sum of the momenta
of the highest momentum track
associated with each vertex of the two Higgs jets (12), 
the largest transverse momentum of an 
identified lepton (electron or muon) with respect to the corresponding 
jet axis (13) and the sum of the two largest numbers of significant tracks in 
the four jets (14).

Finally the likelihood for each event  is required to be larger than
0.90 at $\sqrt{s} = 161$ \Gc\  and larger than 0.925 at $\sqrt{s} = 170-172$
\Gc\  (i.e. ${\cal L}^3(\vec{x}) > \mbox{0.90} ~(\mbox{0.925})$).
The likelihood requirement was tightened for the higher 
center of mass energy in order to retain the same level of
expected background for both energies. 

The analysis was tuned with
a reference mass of
 $\mH = \mbox{65 \Gc\ }( \mbox{68 \Gc\ } )$  
at $\sqrt{s} $= 161 \Gc\  (170-172 \Gc ) 
with a resulting efficiency of   32.1\% (28.5\%). 
The resulting expected background is 0.49 (0.34) events from $Z/\gamma^*$ and 
0.26 (0.55) events from 4-fermion processes. Other sources of background 
are negligible. The total expected background is 0.75$\pm$0.08(stat)$\pm$0.25(syst) 
(0.88$\pm$0.07$\pm$0.18) events. 
The background systematics are dominated by the error 
on the  modeling of the variables
used in the likelihood.

\begin{figure}[htb]
\centerline{\hbox{
\epsfig{figure=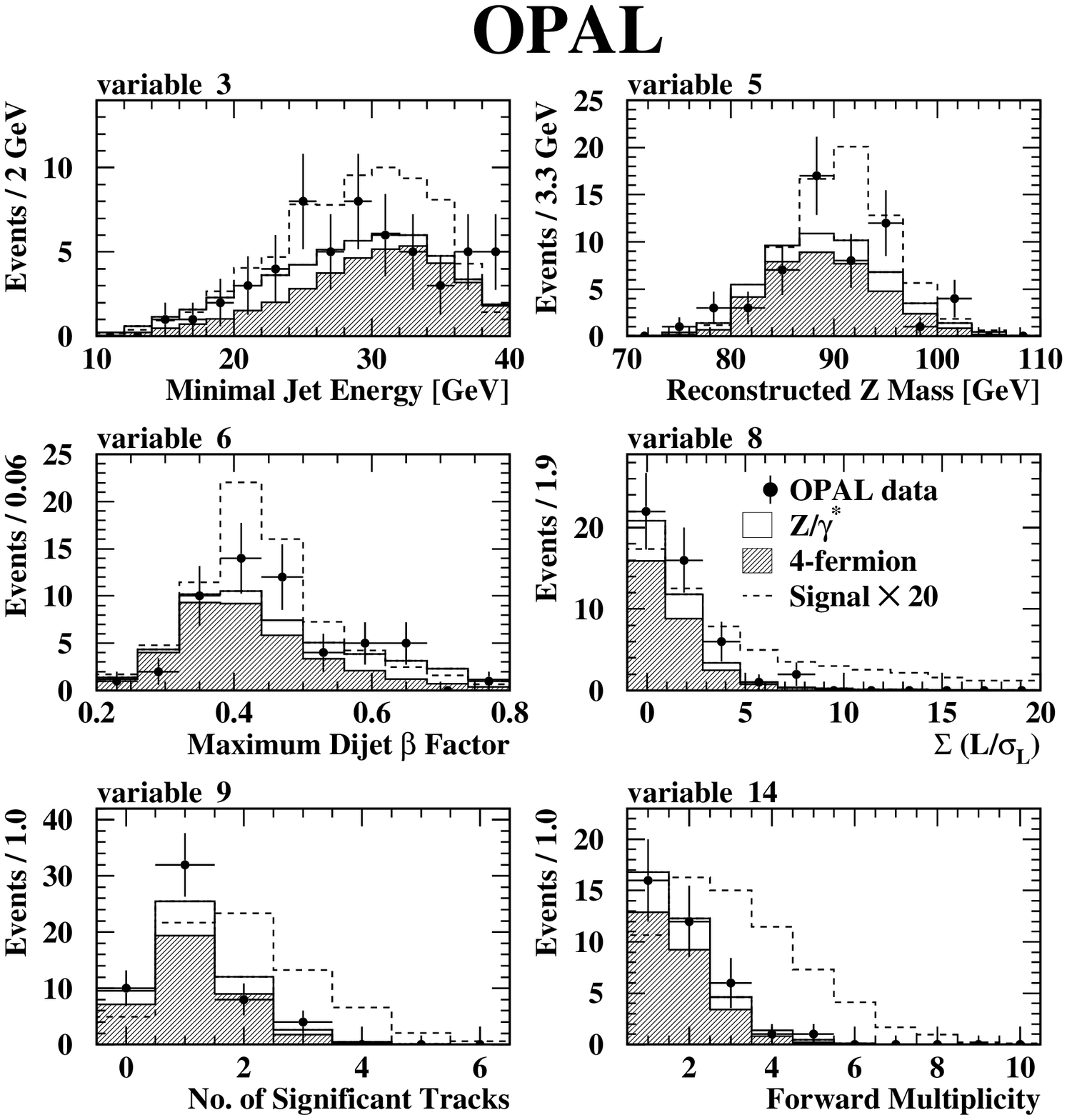,width= 16cm,height=16cm}
}}
\caption[1]
{Four jets channel:
 distribution of input variables 3,5,6,8,9 and 14 (as described 
in the text) for OPAL data at \sqrts=170 and 172 \Gc\ compared
with the Monte Carlo expectations;
 data: points with error bars,
simulations (normalized to
recorded luminosity): 
open / shaded / dashed histograms for \Zo/$\gamma^*$\ra\qq\  /
four-fermion processes / signal (\mH=68~\Gc\@) scaled by a factor of 20.
All distributions are shown after the preselection (cuts 1-6).
}
\label{fig4jets0}
\end{figure}

For all fourteen input quantities a good agreement between OPAL data
 and the Monte Carlo (MC) distributions is
observed. This can be seen in  Figure~\ref{fig4jets0}
where some distributions of data and simulated background as
well as a simulated 68 \Gc\@ Higgs signal
are shown. The agreement  between MC and data,
in particular for b-tag related quantities, was also checked
 at \sqrts=\mZ\ with data taken for calibration immediately before the high
energy runs.

\begin{figure}[htb]
\centerline{\hbox{
\epsfig{figure=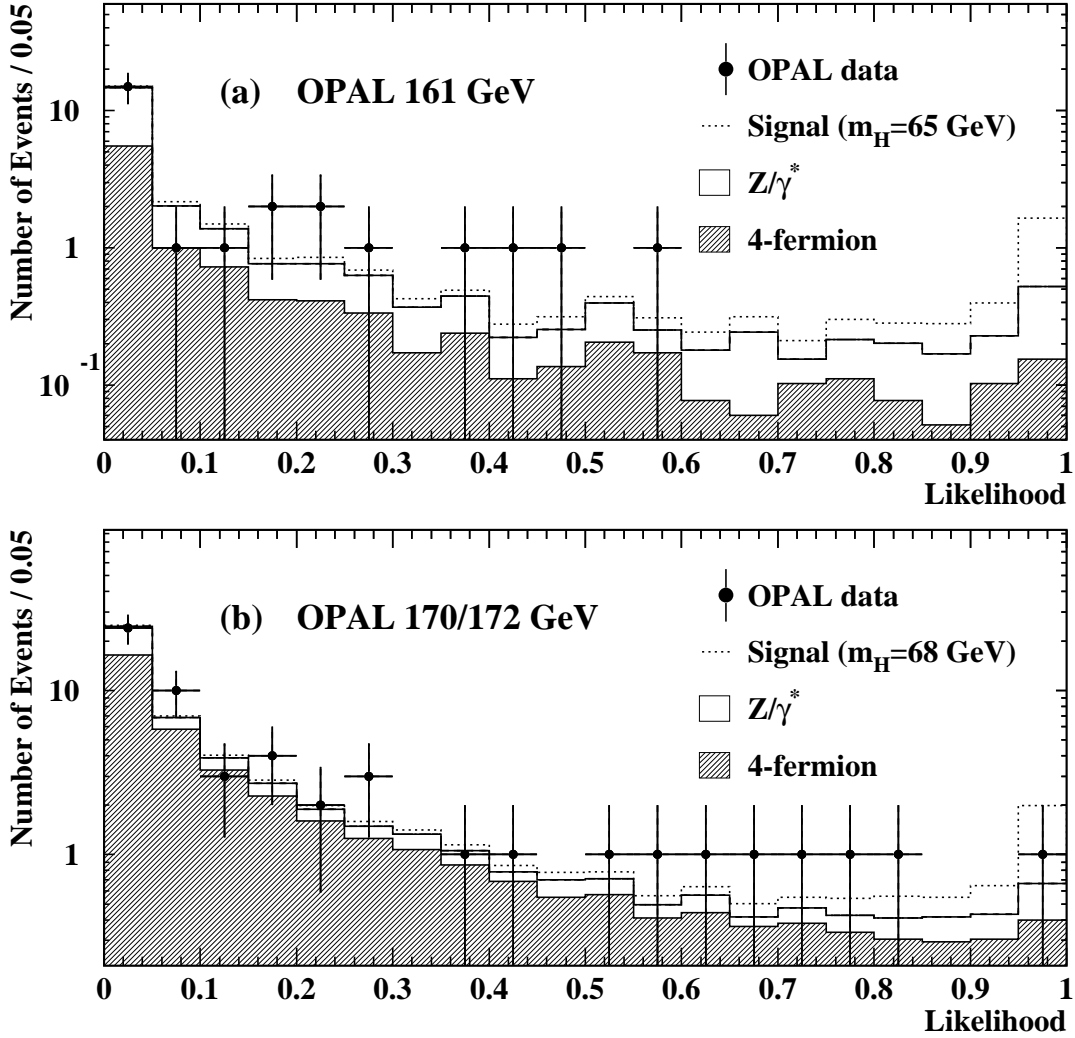,width= 16cm,height=16cm}
}}
\caption[1]
{Four jets channel:
 (a) Likelihood distribution for OPAL data at $\sqrt{s} = 161$~GeV
compared with the Monte Carlo expectation.
(b) Likelihood distribution for OPAL data at
$\sqrt{s} = 170$ and 172~GeV compared with the Monte Carlo expectation.
All distributions are shown after the preselection (cuts 1-6).
}
\label{fig4jets}
\end{figure}

Figures~\ref{fig4jets} (a) and (b) show the distributions of 
the signal likelihood for the preselected events from the 161 \Gc\  and 
170-172 \Gc\  data. The shaded area shows the expectation from  4-fermion
events,
the
white area is $Z/\gamma^*$ events and the dotted line represents the total
background with a signal from a 65 \Gc\  Higgs at $\sqrt{s} = 161$ \Gc\  
and a 68 \Gc\  Higgs at $\sqrt{s} =$ 170-172 \Gc\  added. 
It can be seen that the expected signal is concentrated at large values 
of the likelihood. The peak of the
 background at large
likelihood values is due to irreducible 
signal like four-fermions and QCD  four jets events
containing b flavor.

\begin{figure}[htb]
\centerline{\hbox{
\psfig{figure=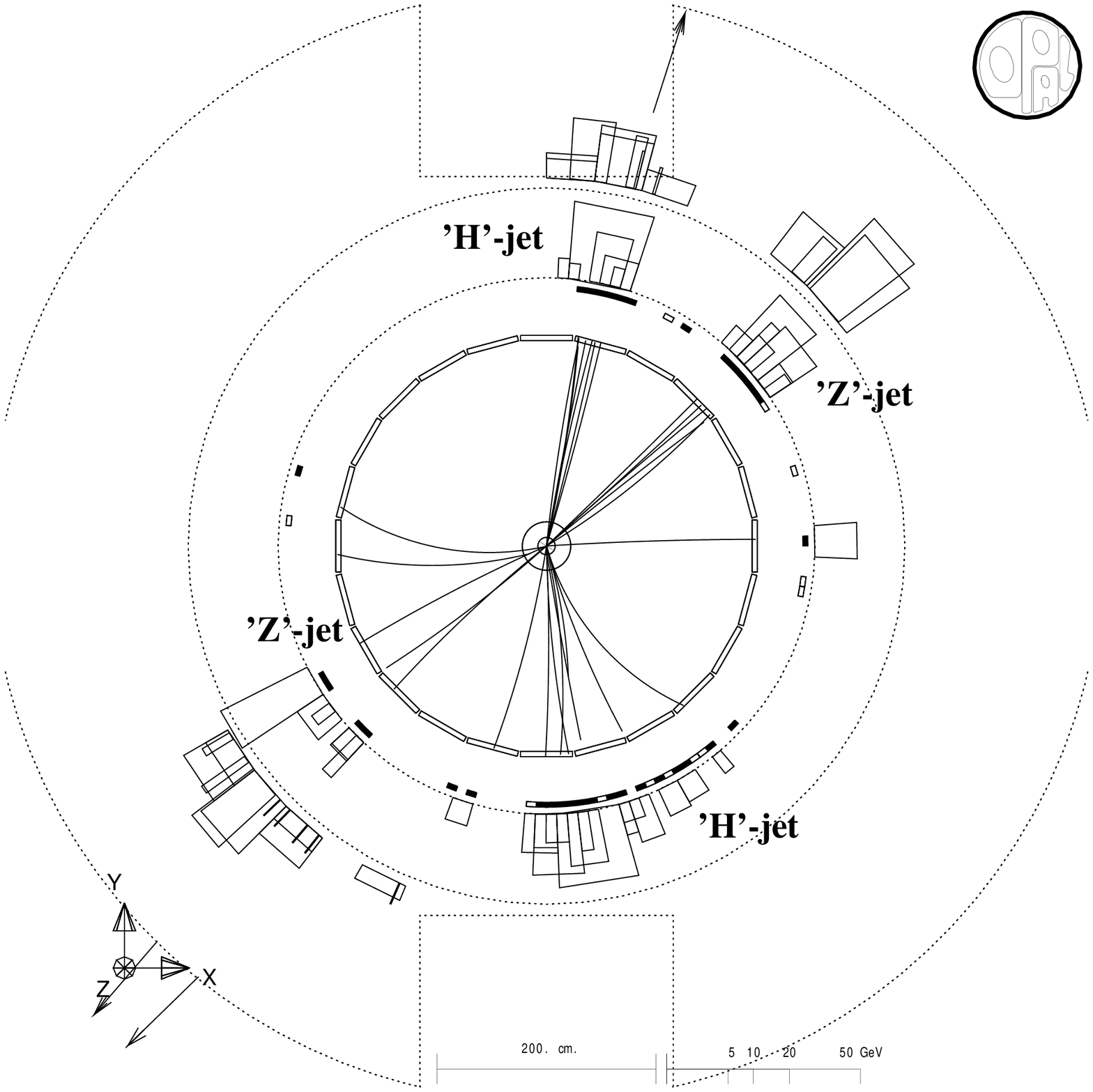,width= 10cm,height=10cm}
}}
\vspace{-2.5cm}
\centerline{\hbox{
\psfig{figure=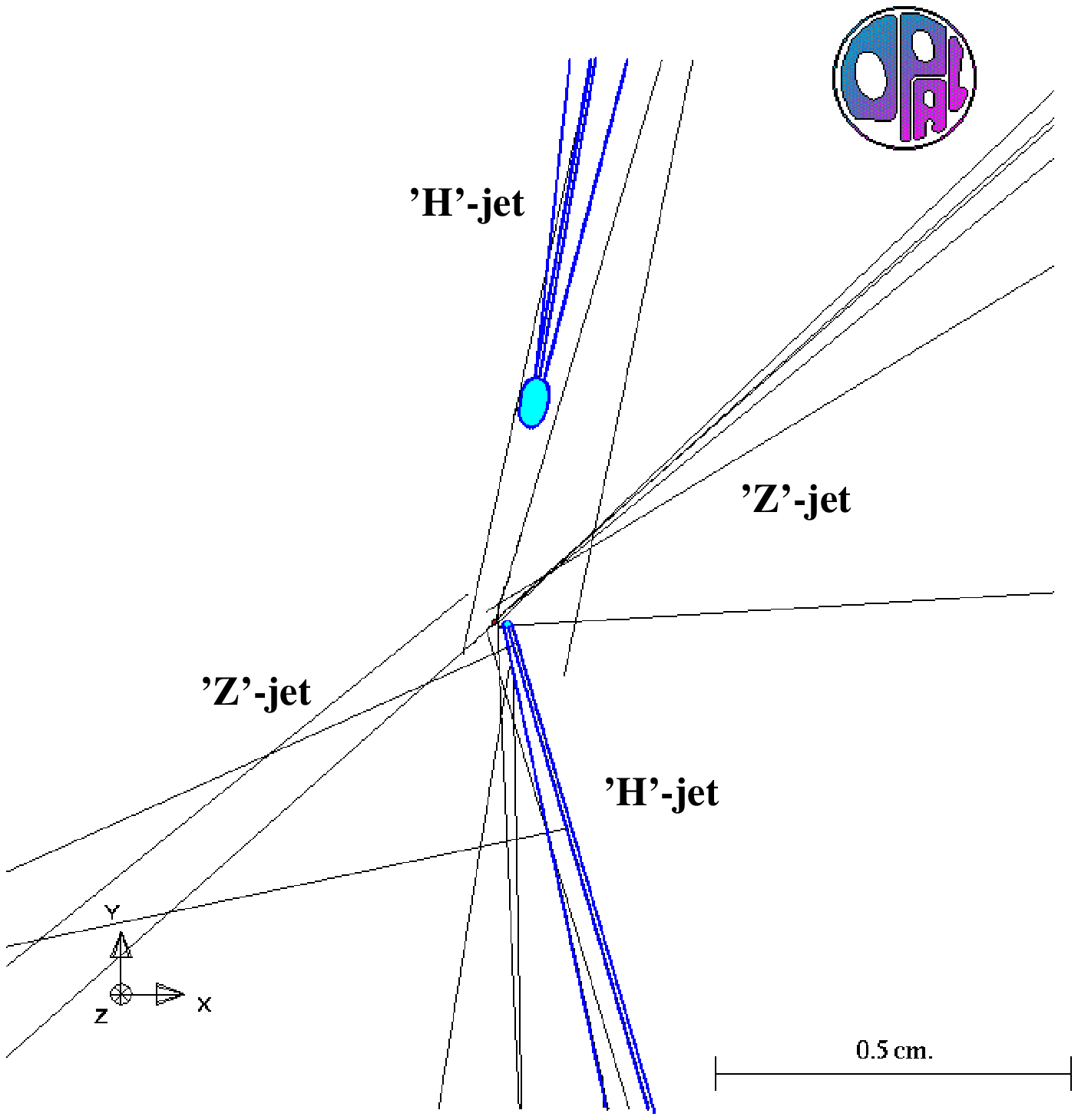,width= 14cm,height=19.5cm}
}}
\vspace{-7cm}
\label{4j_event}
\caption[1]
{The candidate event in the four jet channel. Top: $r\phi$ view with
tracks and clusters; bottom: zoom in on the vertex region. The jets
associated to the Higgs in the kinematic fit yielding the highest
probability, are labeled as Higgs jets ('H'-jet);
 the others are labeled as Z$^{0}$\@ jets ('Z'-jet).
 The error ellipse of the second Higgs jet is too small to be seen
in this plot.
}
\end{figure}

No event survived the likelihood cut of 0.9
at $\sqrt{s}$ = 161 \Gc\@.
With a cut at 0.925 in the likelihood variable one candidate event is selected
from the data collected at $\sqrt{s} =$ 172 \Gc\@.
The likelihood of this event is 0.993. The invariant mass of the two jets
associated with the Higgs boson  decay is $75.6\pm3.0$ \Gc\@. 
Both of the Higgs candidate  jets have displaced secondary vertices with significances 
 of 6.1  and 2.0, 
and charged multiplicities of 4 and 3.
One of the jets assigned to the $Z^0$-candidate has also a displaced 
secondary vertex with 
a significance of 4.6 and multiplicity of 2.
The event is shown in Figure~\ref{4j_event}.

\begin{figure}[htb]
\centerline{\hbox{
\epsfig{figure=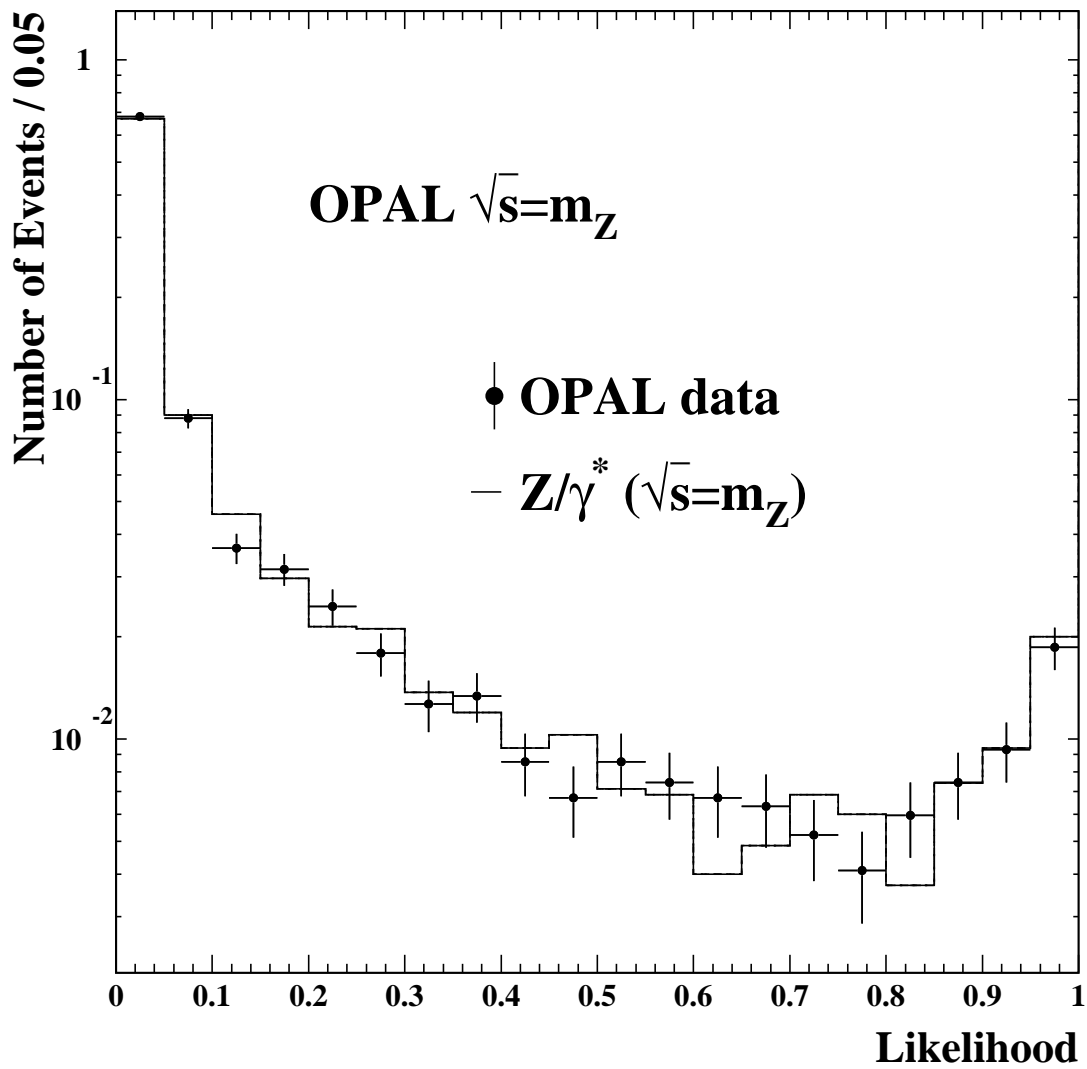,width= 16cm,height=16cm}
}}
\caption[1]
{Four jets channel:
 likelihood distribution of OPAL events at $\sqrt{s} = \mZ\ $
 scaled to LEP 2 energies compared with the Monte Carlo
expectation.
}
\label{fig4jets2}
\end{figure}

The likelihood method has been also
 checked using  calibration events at a
LEP energy of ${\sqrt{s}=m_{\mathrm Z^0}}$ where the physics is well
established.
Events classified as four jets
are selected and 
the tracks momenta and clusters energies are
scaled up to 172~GeV
 center of mass energy.
 The b-tag information and
most of the kinematic event properties are preserved by the procedure.
The likelihood values of the scaled events show a good agreement between
data and MC simulation (see Figure \ref{fig4jets2} ).

 The signal detection efficiencies
at $\sqrt{s}$ = 161 \Gc\@ (170-172 \Gc )  are
affected by the following uncertainties:
Monte Carlo statistical error 4\% (2\%);
description of preselection variables 5\% (2\%);
modeling of the variables used in the likelihood selection, 10\% (7\%).
The latter was dominated by b-tag variables uncertainties
  which were estimated by varying the tracking resolution.
 The modeling was also checked using a re-weighting procedure.
 The difference between data and MC distributions was minimized
by assigning weights to all MC events in an iterative procedure
and recalculating the likelihood with the weighted MC distributions.
The binning of reference distributions yielded a
contribution to the systematic error of 1\% (2\%).
Taking these uncertainties as independent and adding them in
quadrature
results in a total systematic error on the signal efficiency of 12\% (8\%)
relative error.

Two additional analyses were performed
as a  check of the likelihood method.
The first was a cut-based method similar to the published analysis of
the 161~\Gc\  data \cite{opal-161}.
This analysis  gave, for similar
residual background, signal efficiencies which were lower by 10-20\%
than those of the current analysis. Applied to the 170-172 \Gc\  data,
the cut-based analysis did not select the
candidate event selected by the current selection because it  failed
 the cut in the event shape parameter $C$. The $C$-parameter of this event
is 0.546, close to the cut value of 0.55. 

Another multidimensional analysis
utilizing an Artificial Neural Network (ANN) with 
9 b-quark tagging  variables and 12  kinematic variables was
also performed. 
Using the same  Monte Carlo signal and background samples
described previously, the ANN analysis  achieved similar  efficiency, 
signal-to-noise and systematic errors,  to the likelihood approach.
The ANN identified one Higgs boson candidate with a cut on the ANN output
variable corresponding to a background of  0.9 events. This event was the
same as the one identified in the main analysis.

\section{The Missing Energy Channel}
The \ee\ra\nn\Ho\ra\nn\bb\ process amounts to approximately 
18\% of the
Higgs boson signal topologies
with a small contribution (0.7\% for \mH=68~\Gc\@) from the \WW\ fusion process
\ee\ra\nn\Ho\ra\nn\bb\@.
These events are characterized by large missing momentum and two
energetic, acoplanar, hadronic jets containing b-hadrons.
The dominant backgrounds are mismeasured \Zo$/\gamma^*$\ra\qq\ events,
four-fermion processes with neutrinos in the final state, such as
\WW\ra$\ell^{\pm}\nu$\qq\ and W$^{\pm}$e$^{\mp}\nu$\ra\qq~e$^{\mp}\nu$
with the charged lepton escaping detection and, in general, 
events in which 
particles go undetected down the beam pipe such as \ee\ra\Zo$\gamma$
and two-photon events. For most of these backgrounds, the missing momentum
vector points close to the beam direction, while  signal events tend
to have missing momentum in the transverse plane. 

\begin{figure}[htb]
\epsfig{file=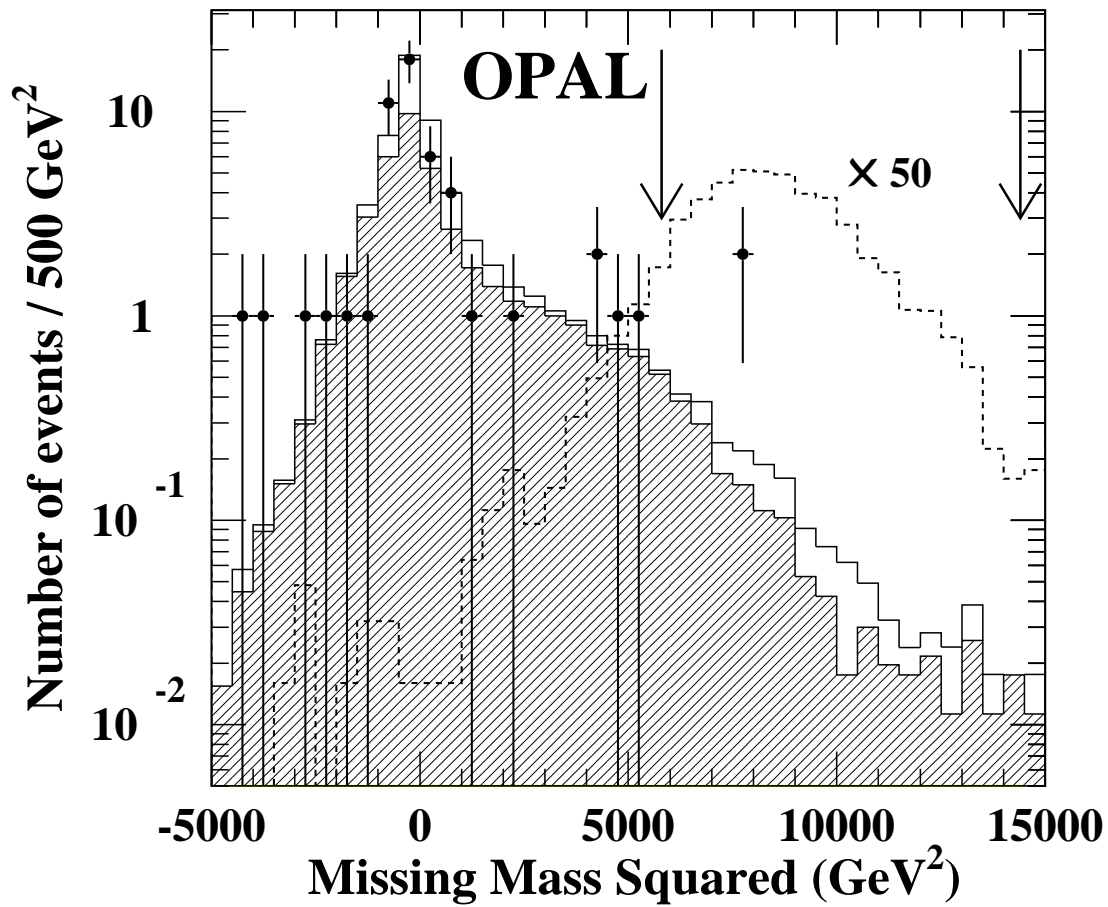,width=0.9\textwidth}
\caption[1]{ 
Missing energy channel; 
 distributions of the missing mass squared after cut (5)
  data: points with error bars,
simulations (normalized to
recorded luminosity): 
open / shaded / dashed histograms for \Zo/$\gamma^*$\ra\qq\  /
four-fermion processes / signal (\mH=68~\Gc\@) scaled by a factor of 50;
The arrows indicate the cut position,
 $76^2 <  m_{\rm miss}^2  < 120^2$~\Gc$^2$.
}
\label{fig_miss1}
\end{figure}

\begin{figure}[htb]
\epsfig{file=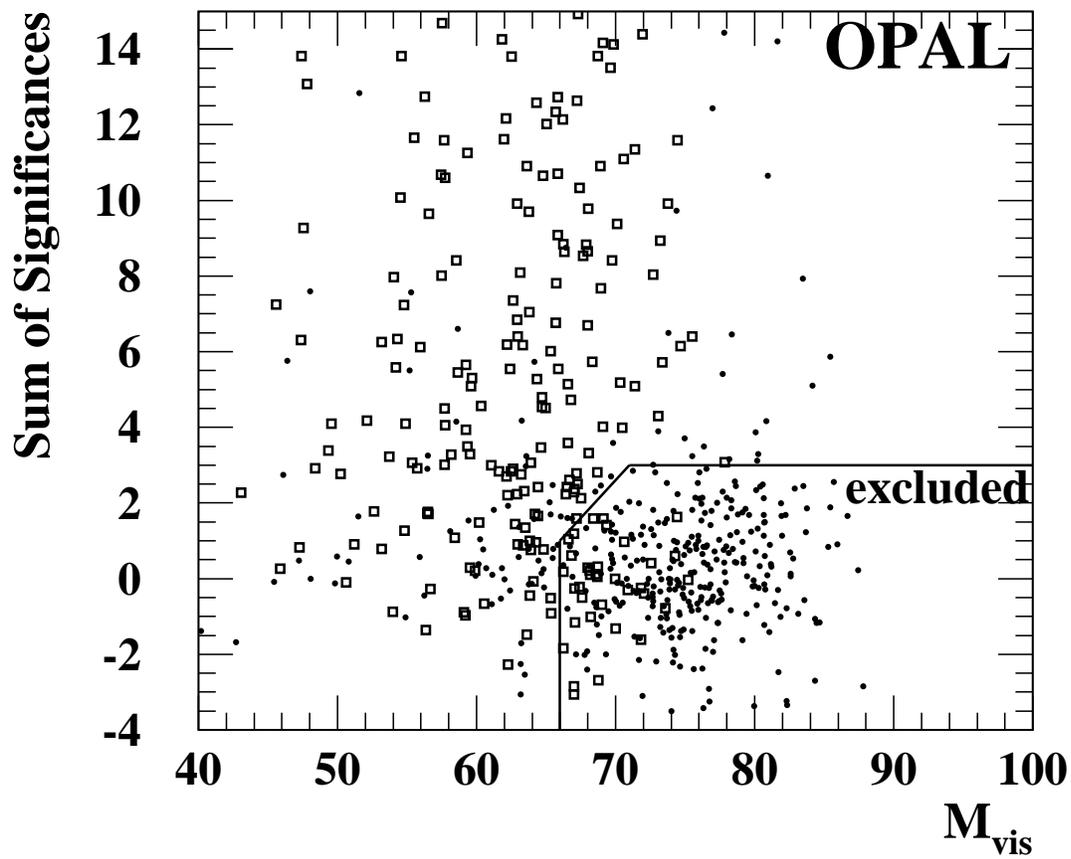,width=0.9\textwidth}
\caption[1]{ 
Missing energy channel; 
  scatter plot of $\Sigma_S$  vs. the visible mass, after cut (7);
dots: simulated four-fermion background; 
open squares: signal for \mH=68~\Gc\@.
 straight line: cut
region. The one event (not shown) which survived the cuts, is at
 a visible mass of 67.7 \Gc\@ and 
  $\Sigma_S$=1.6.
}
\label{fig_miss2}
\end{figure}

The event selection proceeds as follows:
\begin{itemize}
\item[(1)]
To reduce two-photon and beam-wall interactions,
the number of tracks passing the
 quality cuts \cite{higgsold}
should be greater than six, and should
 exceed 20\%
of the total number of tracks in the event.
The energy deposited in the
forward detector, gamma catcher and silicon-tungsten luminometer must be 
less than
2~\Gc\@, 5~\Gc\@, and 5~\Gc\@, respectively. The fraction of energy
deposited in the region 
$|\cos\theta | > 0.9$ must not exceed 50\% of the total visible energy
in the event. The total transverse momentum of the event,
$P^T_{\rm vis}$, must be greater than 1 \Gc\  and the visible mass must
satisfy $m_{\rm vis}> 4$~\Gc\@.

\item[(2)]
To remove backgrounds in which particles go undetected down the
beam pipe, the polar angle, $\theta_{\rm miss}$, of the missing momentum
($\vec{P}_{\rm miss} = -\vec{P}_{\rm vis}$) 
must satisfy $|\cos\theta_{\rm miss}| < 0.9$. The $z$
component of the visible momentum, $P^z_{\rm vis}$, is required to 
be less than 30~\Gc\@.

\item[(3)]
The remaining two-photon background is eliminated by requiring 
$P^T_{\rm vis} > 8$~\Gc\@. As a precaution against large
fluctuations in the measured hadronic energy, $P^T_{\rm vis}$ is recalculated 
excluding hadronic calorimeter clusters and is also required to be larger than
8~\Gc\@.

\item[(4)] The remaining events are reconstructed as two-jet events
  using the Durham algorithm.
   The axes of both jets are required to have a
   polar angle satisfying $|\cos\theta| < 0.9$,
  to ensure good containment.

\item[(5)]  The remaining background from
 \Zo$/\gamma^*$\ra\qq\ is characterized 
 by two jets that tend to be back-to-back
with small acoplanarity angles (where the acoplanarity 
angle is defined as 180$^{\circ} - \phi_{jj}$ 
where $\phi_{jj}$ is the angle between the two jets in the  plane perpendicular to the beam axis),
in contrast to signal events 
in which the jets are expected to 
have some acoplanarity angle due to the boost of the Higgs boson. 
 This background is suppressed by requiring that the
jet-jet acoplanarity angle be larger than $5^{\circ}$. 

\item[(6)]
Since the Higgs boson recoils against a \Zo\ boson decaying into a pair of 
neutrinos, the signal has  a missing mass close to \mZ.
The  remaining backgrounds, predominantly from
well contained multi-hadron and four-fermion events including 
semi-leptonic \WW\ decays, typically have small missing masses.
These backgrounds are reduced by the missing mass requirement  
$76^2 <  m_{\rm miss}^2  < 120^2$~\Gc$^2$.
The distribution of the missing mass squared after cut (5) is shown in Figure 
\ref{fig_miss1}.

\item[(7)]
\WW\ events with one of the W bosons decaying leptonically and the other
 decaying into 
hadronic jets are rejected by requiring that the events have no isolated 
leptons. 

 In this context, leptons are low-multiplicity jets
 with one 
(two or three) tracks associated to electromagnetic or hadronic energy clusters,
confined to a cone of 5$^{\circ}$ (7$^{\circ}$) half-angle,
 having an invariant mass less than 
2.5~\Gc\  and momentum in excess of 5~\Gc\@. The lepton is considered isolated 
if the sum of the track energy and the electromagnetic energy contained
between the above lepton cone and an isolation cone of 25$^{\circ}$ half-angle
does not exceed 10\% (15\%) of the lepton energy. If the lepton cone 
has only one track, the isolation cone is not allowed to contain another track.

\item[(8)] 
The remaining background is mostly from semi-leptonic \WW\ 
and $\mathrm We\nu$ events 
where the charged lepton goes undetected down the beam pipe. 
These events are suppressed
by requiring b-hadrons  in the hadronic jets
of events with a visible mass close to the
$W$ boson mass.
Each of the two hadronic jets is required to contain a secondary vertex
with at least two tracks, each containing two silicon microvertex detector $r\phi$
hits. The two vertices are reconstructed
according to the three dimensional extension of the method used in \cite{btag2}.
The correlation between the sum of the decay length significances,
$\Sigma_S$,
and the visible mass of the event is shown in Figure \ref{fig_miss2} 
where the two-dimensional cut employed is also shown.

\end{itemize}

The numbers of observed and expected events after each selection cut are given 
in Table~\ref{nunu_t1},
along with the detection efficiency for a 68~\Gc\  Higgs boson. 
No events survive the cuts while the expected background is of
0.55$\pm$0.05(stat)$\pm$0.10(syst) events.
The background systematics is dominated by the error 
on the  modelling of the cut variables.

One event is 
rejected only by the b-tag requirement. This event 
  has two well-contained
hadronic jets and evidence for an isolated muon pointing
at the very forward direction where no tracking information
is available. The muon appears as electromagnetic and hadronic clusters aligned with two muon chambers hits.
With a well-measured muon, this
event would be classified as \WW\ra\qq$\mu\nu$ and rejected by  
the  isolated leptons veto.

The detection efficiencies as a function of the Higgs boson mass 
are listed in Table \ref{eff_all}. 
These include a small correction (2.3\%) due to accelerator-related 
backgrounds in the
forward detectors which are not simulated.
The detection efficiencies are affected by
the following uncertainties:
Monte Carlo statistics, 2.2\%;
 modeling of the cut variables other than b-tagging, 3.8\%;
b-tagging and uncertainties from fragmentation
and hadronization
 3.0\%  and lepton tag, 5.6\%.
Taking these uncertainties as independent and adding them in quadrature 
results
in a total relative  systematic uncertainty of 7.7\%.

\begin{table}[htbp]
\begin{center}
\begin{tabular}{|c||r||r||r|r|r||c|} \hline 
 Cut& Data & Total bkg. &\qq($\gamma$)&4-ferm.&\gamgam&Efficiency (\%) \\
    &      &            &             &   &        & \mH~=~68 \Gc\   \\
\hline \hline
(1) &2984 &2829 &713 &105 &2011 & 83.3          \\
(2) &1468 &1486 &302.1 &86.7 &1097 & 73.6         \\
(3) &173 &177.5 &121.0 &56.0 & 0.37 &71.9       \\
(4) &163 &165.9 &113.5 &52.0 &0.30 &64.3       \\
(5) &53 &58.5 &18.3 &40.0 &0.30 &62.3      \\
(6) &2 &2.2 &0.6 & 1.5&0.1&55.2       \\
(7) &1 &1.6 &0.52 &1.0 &0.1 & 52.8\\ 
(8) &0 &0.55$\pm$0.05 &0.275 &0.275 &0 & 42.7\\ 
\hline
\end{tabular}
\end{center}
\caption[sig]{\sl The numbers of events after each cut for the
        data and the expected background for the missing energy channel.
        The background estimates are 
        normalized to 10.4 \pb. The quoted error is statistical.
        The last column shows the selection efficiencies
        for the \nn(\Ho\ra~all) final state,
        for a 68~\Gc\  Higgs boson. }
\label{nunu_t1}
\end{table}

\section{The Tau Channels}
%
The \tptm\qq\ final state can be produced via the processes 
\ee\ra\Zo\Ho\ra\tptm\qq\ (about 3\% of the total \Zo\Ho\ production
rate)
and \qq\tptm (about 5.6\%).
 This analysis is sensitive to both processes, which are
characterized by a pair of tau leptons and a pair of energetic hadronic jets.
In addition, either the pair of hadronic jets or the pair of tau leptons
should have an invariant mass consistent with the \Zo\ mass. These
characteristics are used to suppress the backgrounds, predominantly from 
\Zo$/\gamma^*$\ra\qq\ and four-fermion processes.

The selection begins with the identification of tau leptons using three 
algorithms which address the different
decay channels of the tau lepton.

(a) An electron, identified by a neural
network algorithm~\cite{nn5}, 
 is classified as a 
$\tau^{\pm}$\ra~e$^{\pm}$\nn\ decay if its
momentum is larger than 2~\Gc\@, and it is isolated.
In particular, the number of electromagnetic clusters
within a cone of 26$^\circ$ half-angle around the electron track,
$N_{\rm em}^{26}$, must be less than six, and the ratio of 
the electromagnetic energy within an $11^{\circ}$ cone to that within a
$30^{\circ}$ cone,
$R_{\rm em}^{11/30}$ must be greater than 0.7. There must be no
hadronic calorimeter cluster with energy greater than 0.6~\Gc\  that is
 associated with the electron track. 
Electrons from photon conversions
are rejected using a neural network algorithm \cite{conv}.

(b) A muon, identified using standard selection algorithms~\cite{muon}, 
 is classified as a 
$\tau^{\pm}$\ra$\mu^{\pm}$\nn\ decay if its
momentum is larger than 3~\Gc\@, and if it is isolated.
In particular, $N_{\rm em}^{26} < 5$, and 
the ratio of the scalar sum
of all track momenta within an $11^{\circ}$ cone to that within a
$30^{\circ}$ cone, 
$R_{\rm cd}^{11/30}$,
 must be greater than 0.7.

(c) The remaining tau lepton decays are identified as narrow, isolated
jets. Jets are reconstructed using a 
cone algorithm \cite{cone} with a half-angle of
23$^\circ$ and with at least 3 \Gc\  of associated energy.
Within each resulting jet, the sub-jet of 11$^\circ$ half-angle
having the highest energy is formed.
The sub-jets
are accepted as tau candidates if they satisfy 
the fiducial requirement $|\cos\theta| < 0.92$,
have one or three associated tracks, have an 
invariant mass less than 3.5 \Gc\@, and are isolated 
with $R_{\rm em}^{11/30} > 0.6$. 

In the selection that follows, the tau lepton momentum is approximated
by the momentum of the visible decay products.
When there are two tau lepton candidates 
with momentum vectors separated by less than 23$^\circ$,
one being identified as a leptonic decay (algorithms (a) or (b)) and one as
a narrow jet (algorithm (c)), only the candidate identified as a 
leptonic decay is selected. 

\begin{itemize}
\item[(1)]  
 Events are required to have at least two tau 
lepton candidates, each with
charge of $|q|=1$.

\item[(2)]
  The total track multiplicity of the event must exceed eight. 

\item[(3)]
   Most of the two-photon and \ee\ra\Zo$\gamma$ background events are
   eliminated by requiring that 
 the energy in the forward detector, gamma catcher, and 
  silicon-tungsten luminometer be 
  less  than 4, 10, and 10~\Gc\@, respectively,
  that $|\cos\theta_{\rm miss}| < 0.97$ and that $P^T_{\rm vis} > 3$~\Gc\@.
  In addition, the scalar sum of all track and cluster transverse momenta is
  required to be larger than 40~\Gc\@.
  
\item[(4)]
The remaining \Zo$/\gamma^*$\ra\qq\ background, with and without
photon radiation,
is further suppressed by
requiring that events contain at least four jets, reconstructed using
the cone algorithm with a $23^{\circ}$ half-angle as in (c) above (single 
electrons and muons from tau lepton decays are recognized as 
low-multiplicity ``jets"). 
Events with an energetic isolated photon\footnote{An energetic isolated
photon is defined in this context as an electromagnetic cluster
with energy larger than 15~\Gc\  and no track within a cone
of $30^\circ$ half-angle.} are removed.

\item[(5)]
In  signal events, the algorithms (a), (b), and (c) 
identify 2.3 tau candidates per event on the average. 
Fake tau candidates containing 3 charged tracks are removed by requiring
that the tracks originate from a common vertex in three dimensions, as reflected
by a $\chi^{2}$ probability of a vertex fit larger than 1\%.
Fake candidate pairs are further removed by requiring that the sum of the
track charges
be zero and that the tau candidates satisfy a pairwise isolation requirement,
$|\cos\alpha_1\cdot\cos\alpha_2| < 0.8$,  
  where $\alpha_i$ is the
  angle between the direction of the $i$th tau candidate and that of the
  nearest track not associated with it.
In those rare instances where more than one candidate pair passes the
selection, we give preference to those pairs having taus which were identified
through their leptonic decay mode and, in case of further ambiguity,
we choose the pair with the lowest charged multiplicity and highest isolation.
\end{itemize}
 
The hadronic part of the event, obtained by excluding the 
tracks and clusters from the selected tau candidate pair,
is then split into two jets using the Durham algorithm. The invariant
masses of the tau lepton pair, $m_{\tau\tau}$, and of the  hadron jets,
$m_{\rm had}$, are calculated using only the tau lepton and jet 
momentum directions and 
requiring energy and momentum conservation. At this point the selection
separates into two parts, one (A) sensitive to the 
(\Zo\ra\tptm)(\Ho\ra\qq) process and another (B) sensitive to the
(\Zo\ra\qq)(\Ho\ra\tptm) process.

\begin{itemize}
\item[(6)]
(6A) The selected events must satisfy 
75~\Gc\ $< m_{\tau\tau} < 105$~\Gc\  and  $m_{\rm had}>$30~\Gc\@.
In addition, $E_{\rm vis}$ is required to be less than 155~\Gc\@,
since the neutrinos from the tau lepton decays, originating from the
\Zo\ boson, give rise to a relatively large missing energy. 
Finally,
cuts are implemented to suppress specific four-fermion backgrounds, from
\ee\ra\Zo$/\gamma^*$~+~\Zo$/\gamma^*$ and \ee\ra\Zo\ee. If the tau 
lepton candidates are
both classified as $\tau^{\pm}$\ra~e$^{\pm}$\nn\ or both as 
$\tau^{\pm}$\ra~$\mu^{\pm}$\nn, their 
opening angle is required to be larger than $90^{\circ}$ and, in the first
case, neither electron is allowed to lie within $36^{\circ}$ of the beam axis. 

(6B) The selected events must satisfy 
75~\Gc\ $< m_{\rm had} < 105$~\Gc\  and  $ m_{\tau\tau} > 30$~\Gc\@.
Since in this case the mass cuts are less effective against the background, 
the requirements on the properties of the
tau lepton candidates are tightened. The opening angle of the tau 
lepton pair must be
larger than $110^{\circ}$ and, if one of the tau candidates has a 
track multiplicity exceeding two, the pairwise isolation cut is tightened to
$|\cos\alpha_1\cdot\cos\alpha_2| < 0.55$. Furthermore, to suppress four-fermion
backgrounds, pairs with leptons of the same flavor are rejected.
Finally, to suppress the process \WW\ra$\ell\nu$\qq,
events are rejected if they contain any track or cluster with a
momentum or energy exceeding 40~\Gc\@.
\end{itemize}

\begin{figure}[htb]
\epsfig{file=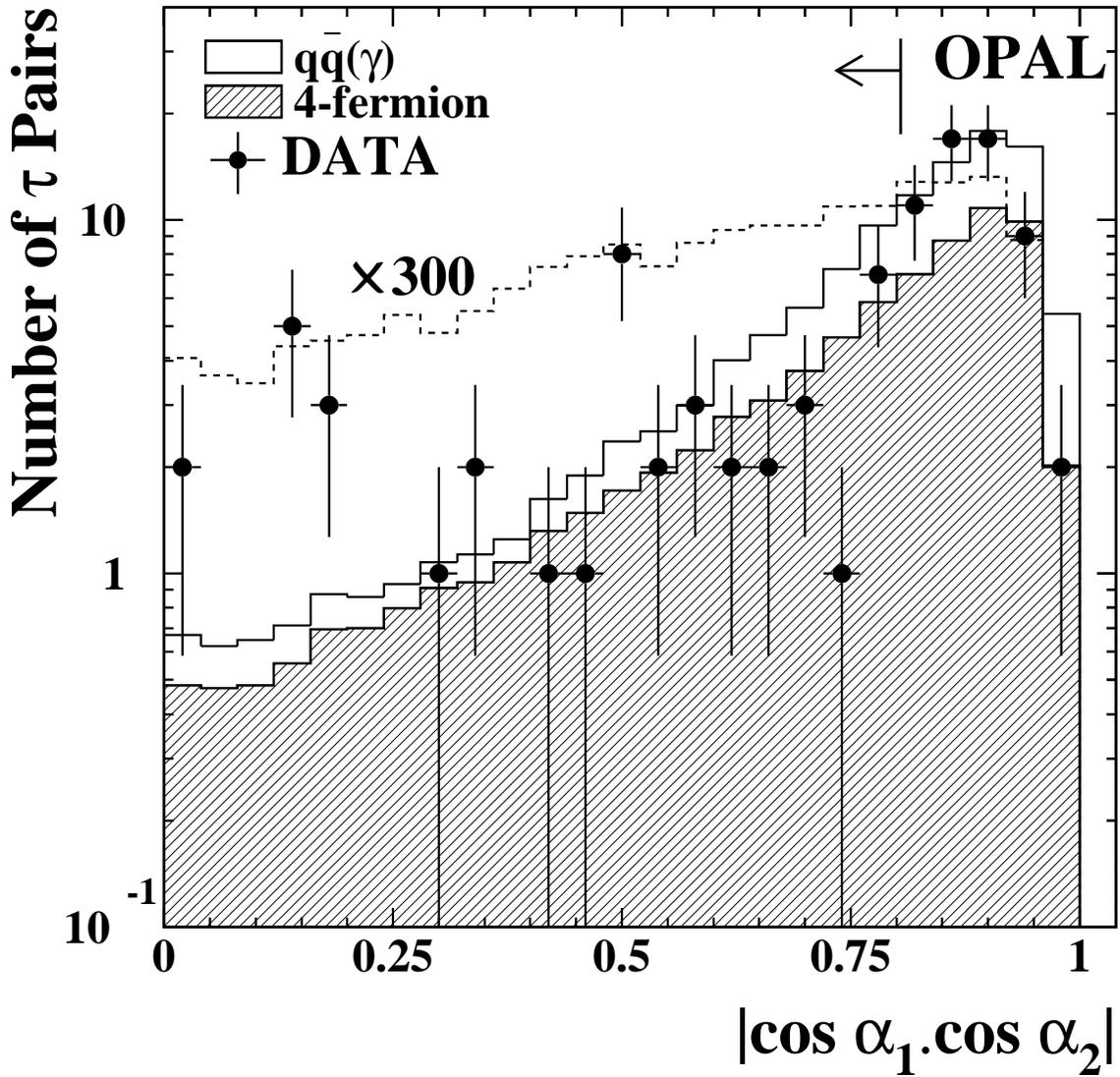,width=0.90\textwidth}
\caption[1]{ 
 Tau channels; distributions of the pairwise
isolation parameter (see text) after cut (3);
 data: points with error bars,
simulations (normalized to
recorded luminosity): 
open / shaded / dashed histograms for \Zo/$\gamma^*$\ra\qq\  /
four-fermion processes / signal (\mH=68~\Gc\ ). Arrows indicate
domains accepted by the cuts.
}
\label{tau1}
\end{figure}

\begin{figure}[htb]
\epsfig{file=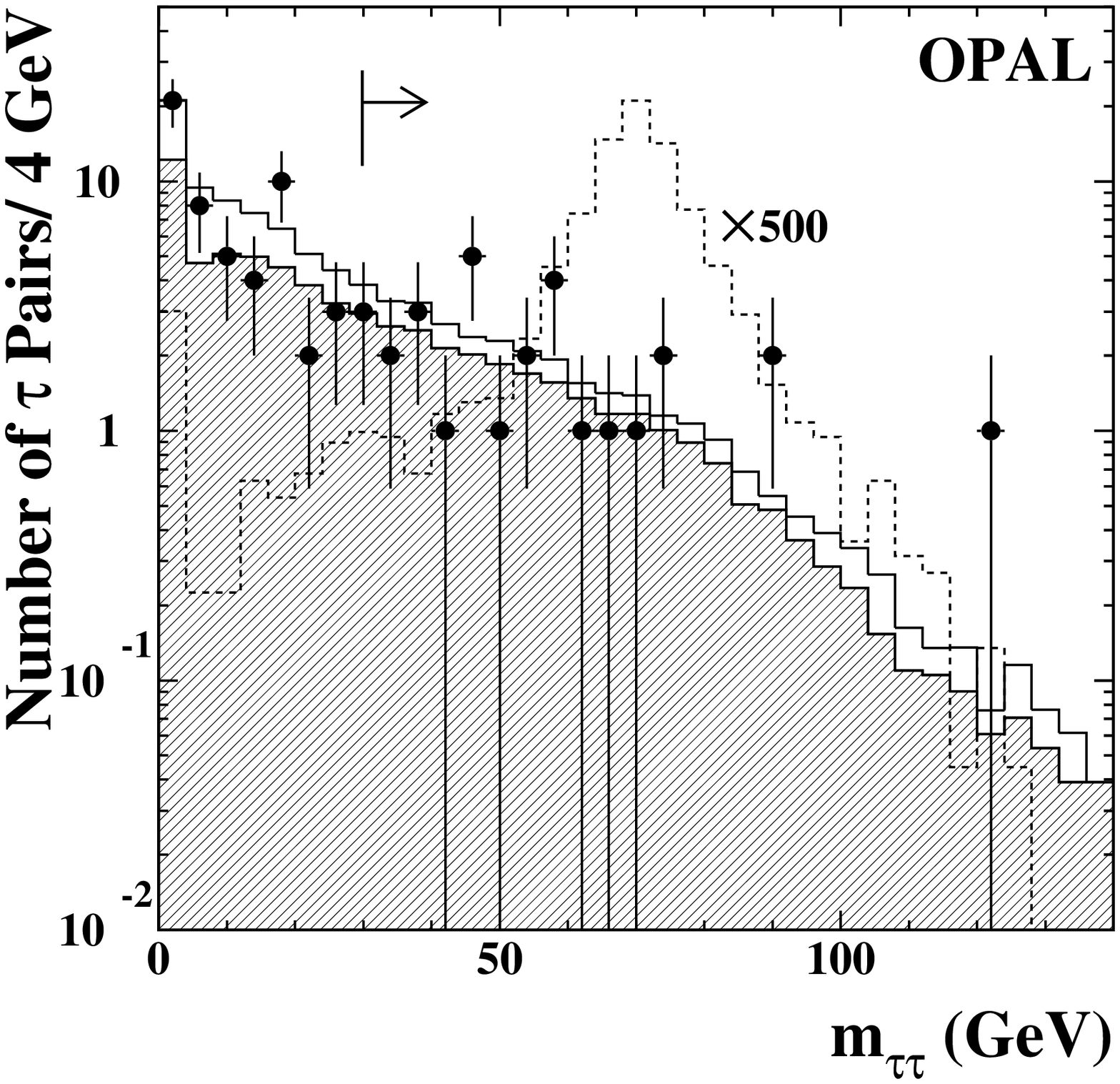,width=0.90\textwidth}
\caption[1]{ 
 Tau channels; distributions of the \tptm\ invariant 
mass after cut (4); 
 data: points with error bars,
simulations (normalized to
recorded luminosity): 
open / shaded / dashed histograms for \Zo/$\gamma^*$\ra\qq\  /
four-fermion processes / signal (\mH=68~\Gc\ ). Arrows indicate
domains accepted by the cuts.
}
\label{tau2}
\end{figure}

Distributions of $|\cos\alpha_1\cdot\cos\alpha_2|$ and $m_{\tau\tau}$ 
are shown in Figures~\ref{tau1}  and \ref{tau2}   for the data, the backgrounds, and
for a 68~\Gc\  Higgs boson signal, with
 \Ho\ra\tptm\@. 
The numbers of observed and expected 
events after each stage of the selection are given in Table \ref{tautau_t1}
where a good agreement between data and MC can be seen.
The detection efficiency for a 68~\Gc\  Higgs boson is also given.
No candidate event is observed while the total background 
for selections A and B is estimated to be
0.59$\pm$0.04(stat)$\pm$0.14(syst) events. 
The background systematics is dominated by the error 
due to the  modeling of the cut variables.

The detection efficiencies 
for cases A and B as a function of the Higgs boson mass are given in
Table \ref{eff_all}.
 and include a small correction coming from 
accelerator-related backgrounds in the
forward detectors (2.3 \%) which are not fully simulated.
The detection efficiencies are affected by the following
uncertainties:
Monte Carlo statistics,  2.8\%;
uncertainty in the tau lepton identification efficiency, 4.3\%;
uncertainties in the modeling of cut variables excluding
the tau lepton identification, 9.1\% (case A) and 7.6\% (case B);
uncertainties in the modeling of fragmentation and hadronization, 1.2\%.
Taking these uncertainties as independent and adding them in quadrature
results in a total systematic uncertainty  of 10.5\% (case A) and 9.2\% 
(case B) (relative errors).
\begin{table}[htbp]
\begin{center}
\begin{tabular}{|c||r||r||r|r|r|r||c|c|} \hline 
 Cut&Data&Total
 bkg.&\qq($\gamma$)&4-ferm.&$\gamma\gamma$&$\ell^+\ell^-$&$\epsilon$(\%), case A
 &$\epsilon$(\%), case B \\
 &    &          &             &       &   &   & 68 \Gc\  & 68 \Gc\  \\
\hline \hline
(1) &$857$ &$611.5$ &$77.8$  &$79.9$ &$421.6$ & $32.2$   &58.5 &59.3 \\
(2) &$358$ &$306.9$ &$75.2$  &$36.8$ &$194.9$ & $0$      &58.3 &58.6 \\
(3) &$50$  &$55.1$  &$23.6$  &$31.2$ &$0.3$   & $0$      &54.0 &52.9 \\
(4) &$37$  &$40.1$  &$15.2$  &$24.7$ &$0.2$   & $0$      &51.7 &50.8 \\
(5) &$15$  &$20.1$  &$6.9$   &$13.2$ &$< 0.07$& $0$      &41.8 &40.8 \\
\hline
(6-A) &$0$   &$0.41\pm 0.03$ &$<0.01$  &$0.41$ &$<0.07$ & $0$ &22.9 &-- \\
\hline
(6-B) &$0$   &$0.18\pm 0.02$ &$0.02$  &$0.16$ &$<0.07$ & $0$ &--   &18.9  \\
\hline
\end{tabular}
\end{center}
\caption[sig]{\sl    The numbers of events after each cut for the
        data and the expected background for the tau channels.
        The background estimate is 
        normalized to 10.4 \pb. The quoted errors are statistical.
        The last two columns show the selection efficiencies, for cases
        A and B, for a 68~\Gc\  Higgs boson. }
\label{tautau_t1}
\end{table}
 \section{The Electron and Muon Channels}
%
The  $\ell^+\ell^-$\qq\ ($\ell =$ e or $\mu$) final states 
arise mainly from the process
\ee\ra\Zo\Ho\ra$\ell^+\ell^-$\qq.
They amount to approximately 
6\% of the
Higgs boson signal topologies
with a small contribution (0.1\% for \mH=68~\Gc\@)
 from the \Zo\Zo\ fusion process
\ee\ra\ee\Ho\ra\ee\qq\@.

The analysis adopted concentrates on those final states proceeding through the
first process. These yield a clean experimental signature in the form of 
large visible energy, two energetic, isolated, oppositely-charged leptons of
the same species reconstructing to the \Zo\ 
boson mass, and two energetic hadronic jets.
The dominant backgrounds are  \Zo$/\gamma^*$\ra\qq\ and four-fermion processes.

The selection proceeds as follows:
\begin{itemize}
\item[(1)]
The selected events are required to have at least six tracks and are 
reconstructed
as four jets, using the Durham algorithm with a cut of $y_{34} >0.001$ (single
electrons or muons are considered as low-multiplicity ``jets''). The events 
must satisfy the relations 
$|p^z_{\rm vis}| < (E_{\rm vis}-0.5\sqrt{s})$   and 
$E_{\rm vis} > 0.6\sqrt{s}$. 
\item[(2)]
The selected events must contain at least one pair of oppositely charged,
same flavor leptons (e or $\mu$).
Muon candidates are identified using standard algorithms~\cite{muon}.
The identification of electron candidates,
optimized for high energy electrons,
 starts by selecting 
electromagnetic calorimeter clusters
having energy larger than 5~\Gc\@ and 90\% of the cluster energy is
deposited in four or five calorimeter cells at most, if the polar angle
of the cluster satisfies 
$|\cos{\theta}|<0.75$ or $|\cos{\theta}|>0.75$, respectively.
If such a cluster is found, tracks are sought within 5.7$^{\circ}$
 of the cluster,
having a momentum larger than 2~\Gc\  and normalized energy-to-momentum 
ratios~\footnote{%
$(E/p)_{\rm norm} = [(E/p)-1]/\sigma$
where $E$ and $p$ are cluster energies and track momenta, and $\sigma$ 
the error associated to $E/p$, obtained from the measurement errors of 
$E$ and $p$.}, $(E/p)_{\rm norm}>-5$.
 If more than one track satisfies 
these conditions, the
one forming the smallest angle with the cluster direction is considered
as the electron track. In the region $|\cos{\theta}|<0.85$, other clusters
within 10$^{\circ}$ of the original one are merged, provided that there is
no other track within 20$^{\circ}$ . 

Muon candidates are grouped into pairs without further restrictions.
For an electron pair to be considered, at least one electron candidate 
must have $-2.5 < (E/p)_{\rm norm} < 5$.

If more than one pair of leptons of the same flavor is found, 
the pair with invariant mass
closest to the \Zo\ boson mass is considered. 

\item[(3)]
Both leptons in the candidate pair must have an energy larger than 20 \Gc\  
with at least one of them larger than 30 \Gc\@. The energy of an electron 
candidate is obtained from the associated electromagnetic clusters 
while for a muon candidate it is approximated by the track momentum.

\item[(4)]
The rest of the event, obtained by excluding the candidate lepton pair,
is reconstructed as two
jets using the Durham algorithm. 
An explicit lepton isolation cut is made to reject the remaining background
from \Zo$/\gamma^*$\ra\qq$(\gamma)$ with real or fake leptons close to
the hadronic jets, 
by requiring that each of the leptons has a 
transverse momentum, calculated with respect to the nearest jet axis, larger 
than 10~\Gc\@. 
The background from \ee\ra\Zo$/\gamma^*$~+~\Zo$/\gamma^*$ is suppressed by
requiring that the opening angle of the jet pair be larger than 50$^\circ$.
\item[(5)]
The selected events must have a lepton pair with an invariant mass
consistent with the \Zo\ boson mass. For electrons the invariant
mass of the lepton pair  must lie between 75 \Gc\  and 105 \Gc\@, 
while for muons it must lie
between 60 \Gc\  and 120 \Gc\@. The differing mass windows take into account the
differing resolutions for electrons and muons.

\end{itemize}

\begin{figure}[htb]
\epsfig{file=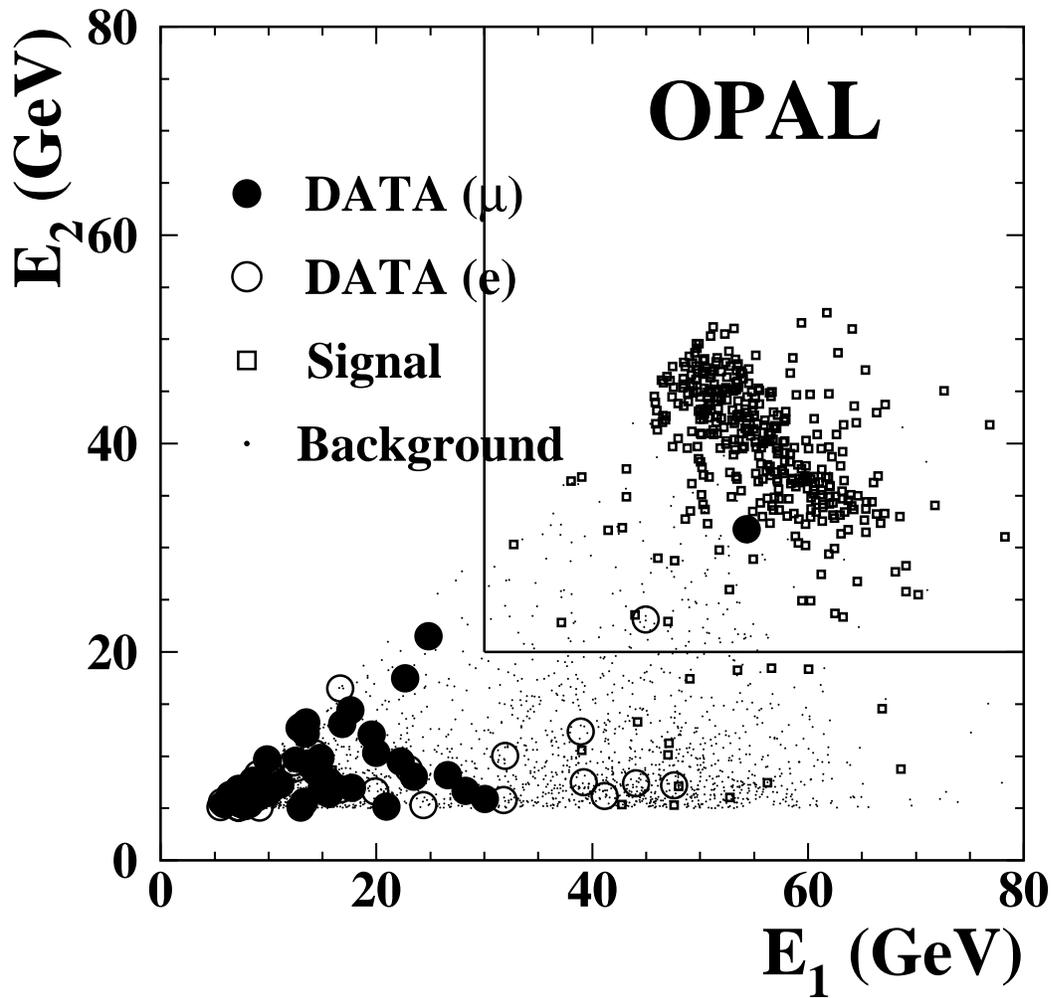,width=0.90\textwidth}
\caption[1]{ 
Electron and muon channels;
 scatter plot of the energies of 
the lepton candidates, after cut (2). The symbols for the data and the simulated
signal (\mH=68~\Gc\@, \ee\ and \mm\ channels combined) as indicated; small dots:
simulated backgrounds. 
Solid lines: position of cut (3).
}
\label{lepton1}
\end{figure}

\begin{figure}[htb]
\epsfig{file=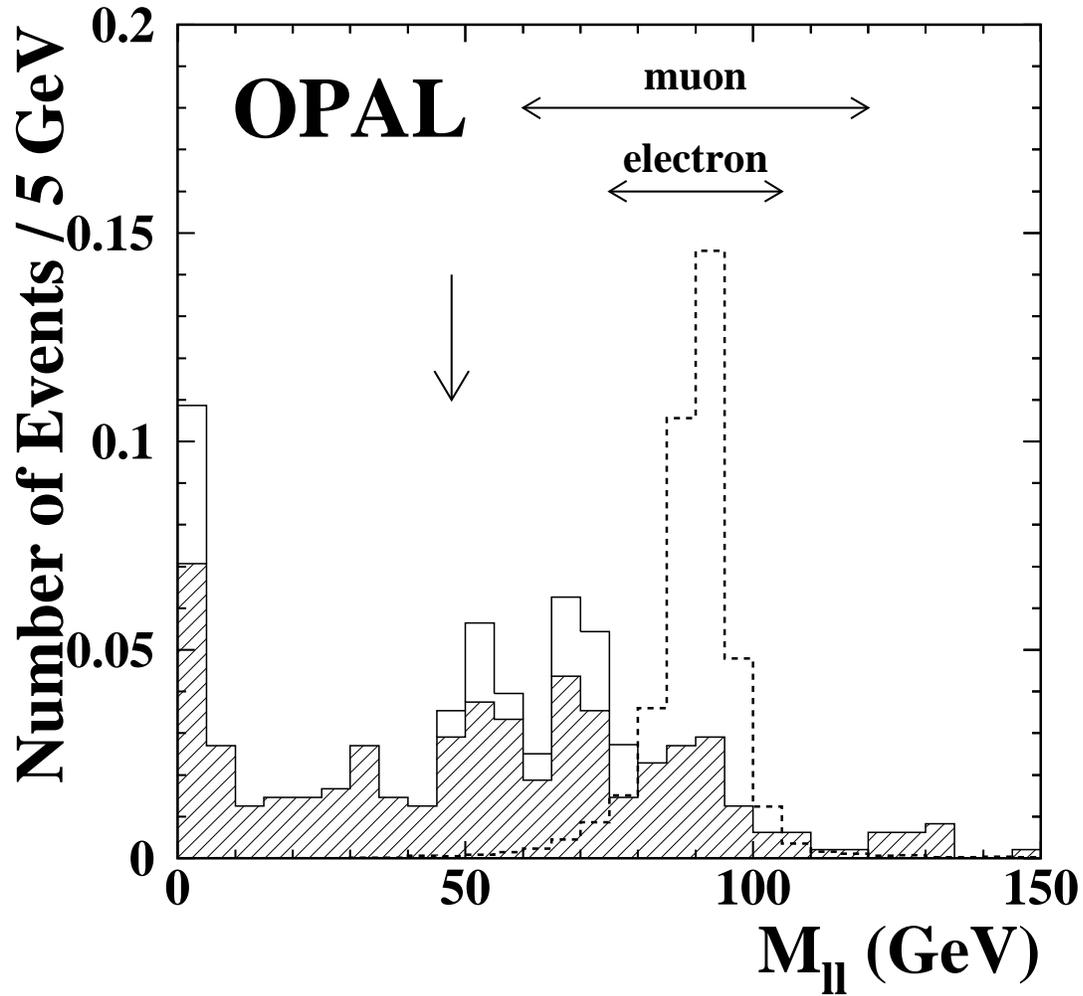,width=0.90\textwidth}
\caption[1]{ 
Electron and muon channels;
 invariant mass of the lepton pair, after cut (4). 
Open / shaded  / dotted histograms:  
\Zo/$\gamma^*$\ra\qq\ / four-fermion backgrounds / simulated signal 
(\mH=68~\Gc\@, \ee\ and \mm\ channels combined). All simulations 
are normalized to the integrated luminosity.
Horizontal arrows: selected mass ranges; 
vertical arrow:  invariant mass of the  event remaining in the \ee\ channel after cut (4).
}
\label{lepton2}
\end{figure}

Distributions of the lepton energy and of the lepton pair invariant mass 
are shown in Figures~\ref{lepton1} and ~\ref{lepton2}, for the data,
the simulated background, and for a simulated signal having \mH=68~\Gc\@. 
The numbers of observed and expected events after each stage of the selection
are given in Table~\ref{tab_lepton}, together with 
the detection efficiency for a 68 \Gc\  Higgs boson. 
After all cuts, no  event survives in either the electron and muon channels,
while in total 0.14$\pm$0.02(stat)$\pm$0.06(syst) events are expected.

The detection efficiencies as a function of the Higgs boson mass are
given in Table \ref{eff_all}.
These are affected by the following systematic uncertainties:
Monte Carlo statistics, 1.0\% (electron), 0.9\% (muon);
uncertainties in the electron (muon) identification, 0.5\% (0.4\%);
uncertainties in the modeling of fragmentation and hadronization, 0.3\%;
uncertainties in modeling the cut variables excluding lepton
identification, 0.4\%.
Taking these uncertainties as independent and adding them in quadrature
results in a total systematic uncertainty of 1.2\% for the electron channel
and 1.1\% for the muon channel (relative errors).
\newcommand{\lw}[1]{\smash{\lower1.7ex\hbox{#1}}}
\begin{table}[htb]
\begin{center}
\begin{tabular}{|r||r||r||r|r||c|c|} \hline
\lw{Cut}&\lw{Data}&\lw{Total bkg.}&\lw{\qq($\gamma$)~~~}&\lw{4-ferm.~~} &
\multicolumn{2}{c|}{Efficiency \mH=68 \Gc\  }\\\cline{6-7}
 & &  & &  & electron (\%)& muon (\%)\\\hline\hline
(1)   & 313 &  335.9 &  246.1 &   89.8 &  88.9 &  82.8 \\
(2)   &  71 &   75.1 &   52.7 &   22.4 &  75.1 &  78.0 \\
(3)   &   2 &    2.0 &    0.9 &    1.1 &  70.5 &  74.9 \\
(4)   &   1 &    0.7 &    0.2 &    0.6 &  67.9 &  72.1 \\
(5)   &   0 &0.14$\pm$0.02&0.02   &0.12&  65.3 &  70.3\\\hline
\end{tabular}
\caption{ \sl The numbers of events after each cut for the
        data and the expected background in the lepton channels.
        Background estimates are normalized to the integrated luminosity. 
        The quoted error is statistical.
        The last two columns show the detection efficiencies for the
        processes \ee\ra(\ee\ or \mm) \Ho\ 
        for a 68 \Gc\  Higgs boson.}
\label{tab_lepton}
\end{center}
\end{table}
\section{Mass Limit for the Standard Model Higgs boson}
%
The signal detection efficiencies 
and the numbers of expected signal events, as a
function of the Higgs boson mass, are
summarized for all search channels in Table~\ref{eff_all}. 

The following uncertainties affecting the numbers of expected signal events
are common to all search channels:
the uncertainty in the integrated luminosity: 0.6\%; 
the uncertainty in the Higgs boson production 
cross section~\cite{gkw}, which includes that from 
the collider energy: 1\%;
and the uncertainty in the Higgs decay branching 
ratios: 2\% \cite{spira,gkw}.
Taking these uncertainties as independent and adding them in quadrature results in a
systematic error, common to all search channels, of 3\% (relative).
In estimating the number of expected events for an assumed Higgs boson mass,
these uncertainties are added in quadrature to those affecting the individual
search channels. 

To derive a new limit on the Higgs boson mass,
this search, with one candidate event in the four jets channel 
at 172~\Gc\@ 
(\mH=$75.6\pm 3.0$~\Gc\@ where the error is
on a Gaussian fit to the peak of a 75\Gc\@ Higgs MC distribution)
 is combined with
earlier OPAL searches at $\sqrt{s}\approx$\mZ~with one candidate
in the leptonic channel (\mH=$61.2\pm 1.0$~\Gc\@)
and 161~\Gc\@  \cite{opal-161}, with 
one candidate in the missing energy channel (\mH=$39.3\pm 4.9$~\Gc\@).
  Two candidates from earlier searches
with \mH$< 25$~\Gc\@  are not considered further.
The expected numbers of Higgs boson events, 
from this search and the combination of  the present and earlier
OPAL searches, are listed in the last two columns of Table~\ref{eff_all}. 
These numbers are affected by total uncertainties of less than 10\%.
\begin{table}[htb]
\begin{center}
\begin{tabular}{|c||c|c|c|c|c|c||c||c|} \hline
\mH\ &\qq\Ho   &\nn\Ho& \tptm\Ho  &\qq\Ho & \ee\Ho &\mm\Ho & $\sqrt s=$170-   
&Grand \\
(\Gc\ )&\Ho\ra\bb&      & \Ho\ra\qq   &\Ho\ra\tptm&    &       &172 \Gc\ 
&total \\
 \hline\hline
 $ 40.0 $&$  14.6 ( 1.7 )$&$  34.0 ( 1.4 )$&$  28.3 ( 0.2 )$&$   2.4 ( 0.0 )$&$  59.6 ( 0.4 )$&$  67.0 ( 0.4 )$&$  4.1 $&$  106.1 $\\
 $ 50.0 $&$  21.7 ( 2.1 )$&$  42.0 ( 1.4 )$&$  28.7 ( 0.1 )$&$   8.3 ( 0.1 )$&$  63.0 ( 0.3 )$&$  71.0 ( 0.4 )$&$  4.4 $&$   34.8 $\\
 $ 55.0 $&$  24.7 ( 2.1 )$&$  46.6 ( 1.3 )$&$  28.0 ( 0.1 )$&$  11.6 ( 0.1 )$&$  63.2 ( 0.3 )$&$  70.1 ( 0.3 )$&$  4.3 $&$   19.6 $\\
 $ 60.0 $&$  27.0 ( 1.9 )$&$  46.9 ( 1.2 )$&$  26.6 ( 0.1 )$&$  15.1 ( 0.1 )$&$  63.6 ( 0.3 )$&$  69.4 ( 0.3 )$&$  3.8 $&$   12.1 $\\
 $ 65.0 $&$  28.2 ( 1.7 )$&$  44.4 ( 0.9 )$&$  24.5 ( 0.1 )$&$  17.9 ( 0.1 )$&$  64.8 ( 0.2 )$&$  69.7 ( 0.2 )$&$  3.2 $&$    6.7 $\\
 $ 67.5 $&$  28.4 ( 1.5 )$&$  43.0 ( 0.8 )$&$  23.2 ( 0.1 )$&$  18.8 ( 0.1 )$&$  65.2 ( 0.2 )$&$  70.2 ( 0.2 )$&$  2.8 $&$    4.7 $\\
 $ 68.0 $&$  28.4 ( 1.5 )$&$  42.7 ( 0.8 )$&$  22.9 ( 0.1 )$&$  18.9 ( 0.1 )$&$  65.3 ( 0.2 )$&$  70.3 ( 0.2 )$&$  2.8 $&$    4.4 $\\
 $ 70.0 $&$  28.2 ( 1.3 )$&$  41.3 ( 0.7 )$&$  21.7 ( 0.1 )$&$  19.3 ( 0.1 )$&$  65.3 ( 0.2 )$&$  70.6 ( 0.2 )$&$  2.4 $&$    3.2 $\\
 $ 75.0 $&$  26.6 ( 0.8 )$&$  34.3 ( 0.4 )$&$  18.2 ( 0.0 )$&$  18.3 ( 0.0 )$&$  63.5 ( 0.1 )$&$  71.3 ( 0.1 )$&$  1.5 $&$    1.6 $\\
 $ 80.0 $&$  23.3 ( 0.2 )$&$  19.2 ( 0.1 )$&$  14.0 ( 0.0 )$&$  14.1 ( 0.0 )$&$  57.6 ( 0.0 )$&$  70.8 ( 0.0 )$&$  0.4 $&$    0.4 $\\
\hline 
\end{tabular}
\end{center}
\caption{ \sl
        Detection efficiencies (in \%) and numbers of expected Higgs 
        boson events 
         (between parentheses) at 172~\Gc\ 
          for each search channel separately 
        as a function of the Higgs boson mass.  The quoted efficiencies
        are obtained from a fit to values determined at fixed values
        of \mH. The last two columns show the total numbers of  
        expected events in the present search at 170 and 172~\Gc\@,
        and the grand total, which also includes the expectations from 
        earlier OPAL searches at center-of-mass energies 
        close to the $Z^0$ mass and the revised  161~\Gc\  analysis.}
 \label{eff_all}
\end{table}

\begin{figure}[htb]
\epsfig{file=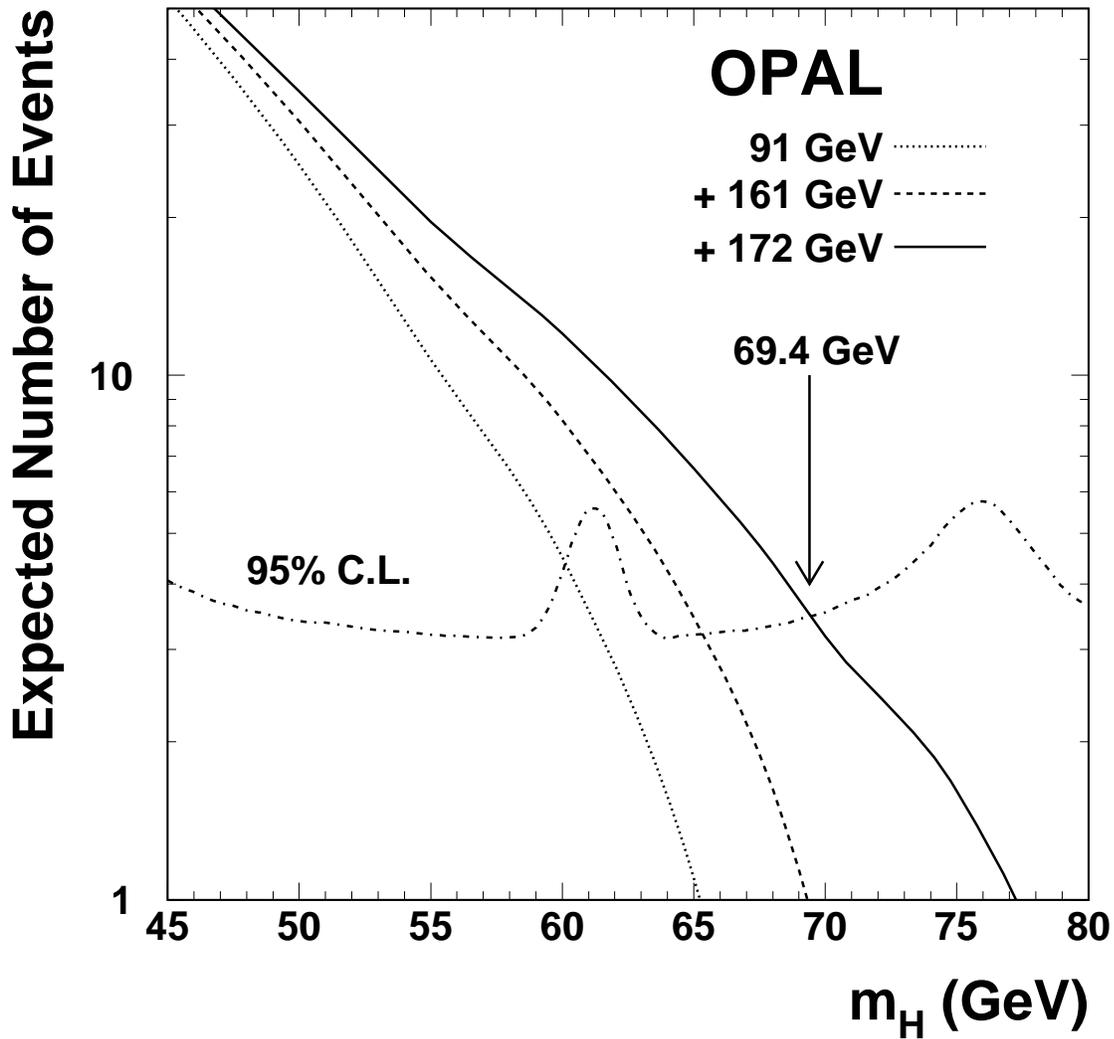,width=0.90\textwidth}
\caption[1]{  
  Expected number of events as a function of the Higgs boson mass. 
 OPAL searches at \sqrts=\mZ\ (dotted), including \sqrts=161~\Gc\  (dashed)
 and the grand total (full line) including OPAL 
searches at \sqrts=170/172~\Gc\@.
The dash-dotted  horizontal  line is the 95\% confidence level upper limit
for a possible Higgs boson signal in the presence of the 
observed high-mass candidate events (see text). 
The intersection of the solid curve with the dash-dotted curve, 
indicated by the
arrow, determines the 95\% confidence level lower limit 
obtained for the Higgs boson mass;
}
\label{limit1}
\end{figure}

Figure~\ref{limit1} shows separately 
the number of expected events for the present search,
for previous OPAL searches 
at the \Zo\ peak and $\sqrt{s}=161$~\Gc\ \cite{opal-161} 
(where the revised 161~\Gc\  analysis is implemented), 
and for their sum, as a function of the Higgs boson mass.
Also shown is  the 95\% confidence level upper limit on the number
of observed candidate events.
 A lower limit on the Higgs boson mass, of 69.4~\Gc\@, is extracted at the
95\% Confidence Level.
 In deriving this limit,
   the probability 
that a candidate event with a given observed mass actually originates 
from a Higgs
boson of arbitrary mass is calculated following Ref.~\cite{bock} which
takes into account the experimental mass resolution including tails 
in the various search channels. 
The expected background is  reduced by the systematic error per channel
and then subtracted. 
It was found that the errors on the background estimation have
marginal effects on the results, reducing the derived limit by
 0.1 \Gc\@. 
The systematic errors are incorporated into the limit according to
the method prescribed in Ref.~\cite{cousins}, reducing the derived limit by
an additional 0.1~GeV. 
The effect of channels weighting and background subtraction is small.
If all channels were assigned  equal weights,
irrespective of the expected rate and background, and no background subtraction had performed, the limit  would have gone down to only 69.0 \Gc\@.

\begin{figure}[htb]
\epsfig{file=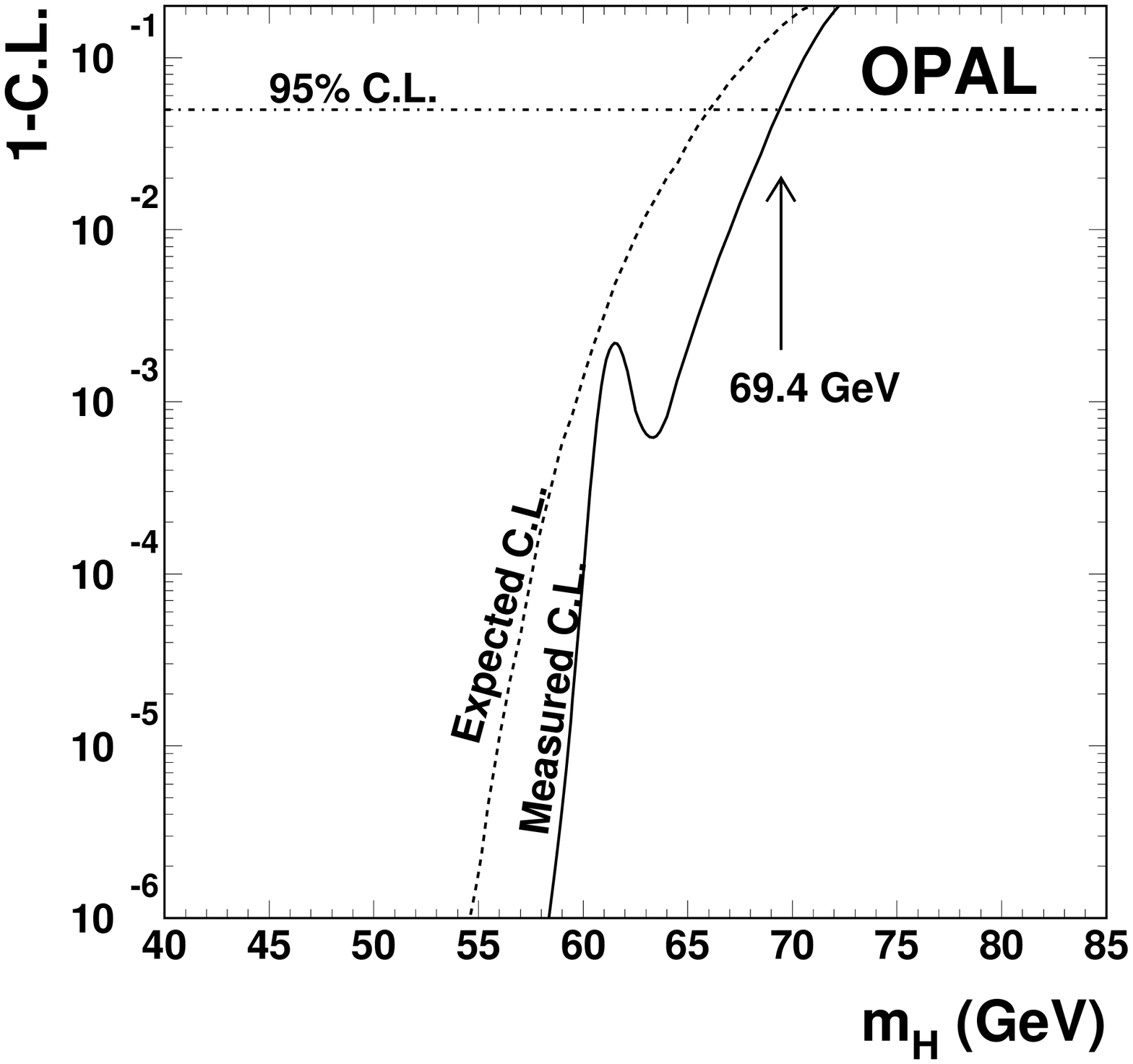,width=0.90\textwidth}
\caption[1]{  
  The expected (dashed) and measured (solid) Confidence Level
 (1-C.L.) as a function of the Higgs boson mass when all
 channels at all center-of-mass energies are combined.
 The intersection of the 95\% C.L. line (dash-dotted) with the dashed
 and solid curves 
  determines the 95\%  expected and measured (respectively) confidence level 
 lower limits 
 obtained for the Higgs boson mass;
 the latter is indicated by an arrow.
}
\label{limit2}
\end{figure}

Figure~\ref{limit2} shows the measured confidence level and the expected one
(averaged over a large number of hypothetical experiments with
no signal and candidates spread according to the
expected background distributions) as a function of the Higgs boson mass.
From this Figure it can be seen that 
the  expected limit is at 65.4~\Gc\ 
and the  expected C.L. for the experimental limit of
69.4~\Gc\  is  82\%.
According to MC trial experiments
a reasonable value of 11.8\% was found for 
the probability of inferring a mass limit greater
    than or equal to 69.4 \Gc\
   assuming no contribution from a Higgs boson
    signal.


%
\section{Summary}
A new search is presented for the Standard Model Higgs boson produced 
in association
with a fermion-antifermion pair.
The search is 
based on data collected in 1996
by the OPAL experiment at center-of-mass energies of 170 and 172~\Gc\@,
with an integrated luminosity
of 10.4~\pb. The data show no significant excess beyond the
background predicted by the Standard Model. 
Combined with earlier OPAL searches at center-of-mass
energies in the vicinity of the \Zo\ resonance and 
a revised analysis of the 161~\Gc\ data,
this search leads to a lower limit of 69.4~\Gc\ 
for the mass of the Standard Model Higgs boson, at the 95\% confidence level.
\bigskip

\pagestyle{empty}
\newpage



\begin{thebibliography}{99}
%
\bibitem{sm}
   S. L. Glashow, J. Iliopoulos and L. Maiani, Phys. Rev.
   {\bf D2} (1970) 1285;\\
   S. Weinberg, Phys. Rev. Lett. {\bf 19} (1967) 1264;\\
   A. Salam, {\it Elementary Particle Theory}, ed. N. Svartholm
   (Almquist and Wiksells, Stockholm, 1968), 367.
%
\bibitem{higgs}
   P. W. Higgs, Phys. Lett. {\bf 12} (1964) 132;\\
   F. Englert and R. Brout, Phys. Rev. Lett. {\bf 13} (1964) 321;\\
   G. S. Guralnik, C. R. Hagen, and T. W. B. Kibble,
   Phys. Rev. Lett. {\bf 13} (1964) 585.
%
\bibitem{sm-all}
ALEPH Collaboration, D. Buskulic \etal, Phys.~Lett.~{\bf B384} (1996) 427;\\
DELPHI Collaboration, P. Abreu \etal, Nucl. Phys. {\bf B421} (1994) 3;\\
L3 Collaboration, M. Acciarri \etal, Phys.~Lett.~{\bf B385} (1996) 454;\\
OPAL Collaboration, G.~Alexander \etal, Z.~Phys.~{\bf C73} (1997) 189.
%
\bibitem{opal-161}
OPAL Collaboration, K. Ackerstaff \etal, CERN-PPE/96-161, 
Phys. Lett. {\bf B393} (1997) 231.

%
\bibitem{spira}
A. Djouadi, M. Spira, and P. M. Zerwas, Z. Phys. {\bf C70} (1996) 425.
%
\bibitem{detector}
OPAL Collaboration, K.~Ahmet \etal, Nucl. Inst. and Meth.
{\bf A305} (1991) 275.
%
\bibitem{simvtx}
P. P. Allport \etal, Nucl. Inst. and Meth. {\bf A346} (1994) 476;\\
S. Anderson \etal, submitted to Nucl. Inst. and Meth. {\bf A}.
%
\bibitem{sw}
B.E. Anderson \etal, IEEE Transactions on Nuclear Science {\bf 41}
(1994) 845.
%
\bibitem{lumino}
OPAL Collaboration, K. Ackerstaff \etal, Phys. Lett. {\bf B391} (1997) 221.
%
\bibitem{higgsold}
OPAL Collaboration, R. Akers \etal, Phys. Lett. {\bf B327} (1994) 397.
%
\bibitem{gopal}
J. Allison \etal, Nucl. Inst. and Meth. {\bf A317} (1992) 47.
%
\bibitem{hzha}
HZHA generator:
P. Janot, 
in {\it Physics at LEP2}, edited by G. Altarelli, T. Sj\"ostrand and
F. Zwirner, CERN 96-01, vol. 2 (1996), p. 309. For Higgs production
cross sections see also \cite{gkw}. 
%
\bibitem{gkw}
E. Gross, B. A. Kniehl, and G. Wolf, Z. Phys. {\bf C63} (1994) 417; \\
erratum {\it ibid.} {\bf C66} (1995) 32.
%
\bibitem{pythia}
PYTHIA 5.721 and JETSET 7.408 generators:
T. Sj\"ostrand, Comp. Phys. Comm. {\bf 82} (1994) 74; 
T. Sj\"ostrand, LU TP 95-20.
%
\bibitem{grc4f}
The grc4f 1.1 generator:
J. Fujimoto \etal, Comput. Phys. Commun. 100 (1997) 128.\\
J.~Fujimoto \etal, in {\it Physics at LEP2},
edited by G.~Altarelli, T.~Sj\"{o}strand and
F.~Zwirner, CERN 96-01, vol.~2 (1996) p. 30.
%
\bibitem{bhwide}
BHWIDE generator:
S.~Jadach, W.~P{\l}aczek, B.F.L.~Ward,
in {\it Physics at LEP2},
edited by G.~Altarelli, T.~Sj\"{o}strand and
F.~Zwirner, CERN 96-01, vol. 2 (1996), p. 286; UTHEP-95-1001.
%
\bibitem{koralz}
KORALZ 4.0 generator:
S.~Jadach, B.~F.~L.~Ward, Z.~W\c{a}s, Comp. Phys. Comm. {\bf 79} (1994) 503.
%
\bibitem{phojet}
PHOJET 1.05 generator: E. Budinov \etal, in {\it Physics at LEP2},
edited by G.~Altarelli, T.~Sj\"{o}strand and
F.~Zwirner, CERN 96-01, vol.~2 (1996) p. 216; \\
R. Engel and J. Ranft, Phys. Rev. {\bf D54} (1996) 4244.
%
\bibitem{vermaseren}
J.A.M.~Vermaseren, Nucl. Phys. {\bf B229} (1983) 347.
%
\bibitem{tkmh}
OPAL Collaboration, G. Alexander \etal, Z. Phys. {\bf C52} (1991) 175.
%
\bibitem{sprim}
OPAL Collaboration, G. Alexander \etal, Phys. Lett. {\bf B376} (1996) 232.
%
%
\bibitem{durham}
N. Brown and W. J. Stirling, Phys. Lett. {\bf B252} (1990) 657; \\
S. Bethke, Z. Kunszt, D. Soper and W. J. Stirling, Nucl. Phys. {\bf B370}
(1992) 310; \\
S. Catani \etal, Phys. Lett. {\bf B269} (1991) 432; \\
N. Brown and W. J. Stirling, Z. Phys. {\bf C53} (1992) 629.
%
\bibitem{cpar}
G. Parisi, Phys. Lett. {\bf B74} (1978) 65; \\
J. F. Donoghue, F. E. Low and S. Y. Pi, Phys. Rev. {\bf D20} (1979) 2759.
%
%
\bibitem{btag2}
OPAL Collaboration, R. Akers \etal, Z. Phys. {\bf C65} (1995) 17.
%
\bibitem{nn5}
OPAL Collaboration, R.~Akers \etal, Phys. Lett. {\bf B327} (1994) 411.
%
\bibitem{conv}
OPAL Collaboration, G. Alexander \etal, Z. Phys. {\bf C70} (1996) 357.
%
\bibitem{muon}
OPAL Collaboration, G. Alexander \etal, Z. Phys. {\bf C52} (1991) 175.
%
%
\bibitem{cone} 
OPAL Collaboration, R. Akers \etal, Z. Phys {\bf C63} (1994) 197.
\bibitem{bock}
P. Bock, {\it Determination of exclusion limits for particle production
using different decay channels with different efficiencies, mass
resolutions and backgrounds}, Heidelberg University
preprint HD-PY-96/05 (1996).
%
\bibitem{cousins}
R. D. Cousins, V. L. Highland, Nucl. Inst. and Meth. {\bf A320} (1992) 331.
%
\end{thebibliography}
 \end{document}